\documentclass[twocolumn,twocolappendix]{aastex631}
\usepackage{xcolor}
\usepackage{CJK}
\usepackage[utf8]{inputenc}

\newcommand{\msun}{M\ensuremath{_{\odot}}}

\newcommand{\rsun}[1]{$\mathrm{R_{\odot}}$}
\newcommand{\lsun}[1]{$\mathrm{L_{\odot}}$}
\newcommand{\Halpha}[1]{$\mathrm{H\alpha}$}
\newcommand{\Hbeta}[1]{$\mathrm{H\beta}$}

\begin{document}

\title{Extended Shock Breakout and Early Circumstellar Interaction in SN 2024ggi}

\correspondingauthor{M. Shrestha}
\email{mshrestha1@arizona.edu}

\newcommand{\LCO}{\affiliation{Las Cumbres Observatory, 6740 Cortona Drive, Suite 102, Goleta, CA 93117-5575, USA}}
\newcommand{\UCSB}{\affiliation{Department of Physics, University of California, Santa Barbara, CA 93106-9530, USA}}
\newcommand{\KITP}{\affiliation{Kavli Institute for Theoretical Physics, University of California, Santa Barbara, CA 93106-4030, USA}}
\newcommand{\UCD}{\affiliation{Department of Physics and Astronomy, University of California, Davis, 1 Shields Avenue, Davis, CA 95616-5270, USA}}
\newcommand{\WIS}{\affiliation{Department of Particle Physics and Astrophysics, Weizmann Institute of Science, 76100 Rehovot, Israel}}
\newcommand{\OKC}{\affiliation{Oskar Klein Centre, Department of Astronomy, Stockholm University, Albanova University Centre, SE-106 91 Stockholm, Sweden}}
\newcommand{\OAPD}{\affiliation{INAF-Osservatorio Astronomico di Padova, Vicolo dell'Osservatorio 5, I-35122 Padova, Italy}}
\newcommand{\Caltech}{\affiliation{Cahill Center for Astronomy and Astrophysics, California Institute of Technology, Mail Code 249-17, Pasadena, CA 91125, USA}}
\newcommand{\GSFC}{\affiliation{Astrophysics Science Division, NASA Goddard Space Flight Center, Mail Code 661, Greenbelt, MD 20771, USA}}
\newcommand{\UMD}{\affiliation{Joint Space-Science Institute, University of Maryland, College Park, MD 20742, USA}}
\newcommand{\UCB}{\affiliation{Department of Astronomy, University of California, Berkeley, CA 94720-3411, USA}}
\newcommand{\TTU}{\affiliation{Department of Physics, Texas Tech University, Box 41051, Lubbock, TX 79409-1051, USA}}
\newcommand{\STScI}{\affiliation{Space Telescope Science Institute, 3700 San Martin Drive, Baltimore, MD 21218-2410, USA}}
\newcommand{\UT}{\affiliation{University of Texas at Austin, 1 University Station C1400, Austin, TX 78712-0259, USA}}
\newcommand{\IoA}{\affiliation{Institute of Astronomy, University of Cambridge, Madingley Road, Cambridge CB3 0HA, UK}}
\newcommand{\QUB}{\affiliation{Astrophysics Research Centre, School of Mathematics and Physics, Queen's University Belfast, Belfast BT7 1NN, UK}}
\newcommand{\IPACSSC}{\affiliation{Spitzer Science Center, California Institute of Technology, Pasadena, CA 91125, USA}}
\newcommand{\IPAC}{\affiliation{IPAC, California Institute of Technology, 1200 East California Boulevard, Pasadena, CA 91125, USA}}
\newcommand{\JPL}{\affiliation{Jet Propulsion Laboratory, California Institute of Technology, 4800 Oak Grove Dr, Pasadena, CA 91109, USA}}
\newcommand{\Southampton}{\affiliation{Department of Physics and Astronomy, University of Southampton, Southampton SO17 1BJ, UK}}
\newcommand{\LANL}{\affiliation{Space and Remote Sensing, MS B244, Los Alamos National Laboratory, Los Alamos, NM 87545, USA}}
\newcommand{\Tsinghua}{\affiliation{Physics Department and Tsinghua Center for Astrophysics, Tsinghua University, Beijing, 100084, People's Republic of China}}
\newcommand{\NAOC}{\affiliation{National Astronomical Observatory of China, Chinese Academy of Sciences, Beijing, 100012, People's Republic of China}}
\newcommand{\YNAO}{\affiliation{Yunnan Observatories (YNAO), Chinese Academy of Sciences (CAS), Kunming, 650216, People's Republic of China}}
\newcommand{\ICEY}{\affiliation{International Centre of Supernovae, Yunnan Key Laboratory, Kunming 650216, People's Republic of China}}
\newcommand{\Itagaki}{\affiliation{Itagaki Astronomical Observatory, Yamagata 990-2492, Japan}}
\newcommand{\Einstein}{\altaffiliation{Einstein Fellow}}
\newcommand{\Hubble}{\altaffiliation{Hubble Fellow}}
\newcommand{\CfA}{\affiliation{Center for Astrophysics \textbar{} Harvard \& Smithsonian, 60 Garden Street, Cambridge, MA 02138-1516, USA}}
\newcommand{\UA}{\affiliation{Steward Observatory, University of Arizona, 933 North Cherry Avenue, Tucson, AZ 85721-0065, USA}}
\newcommand{\MPIA}{\affiliation{Max-Planck-Institut f\"ur Astrophysik, Karl-Schwarzschild-Stra\ss{}e 1, D-85748 Garching, Germany}}
\newcommand{\DSFP}{\altaffiliation{LSSTC Data Science Fellow}}
\newcommand{\HCO}{\affiliation{Harvard College Observatory, 60 Garden Street, Cambridge, MA 02138-1516, USA}}
\newcommand{\Carnegie}{\affiliation{Observatories of the Carnegie Institute for Science, 813 Santa Barbara Street, Pasadena, CA 91101-1232, USA}}
\newcommand{\TAU}{\affiliation{School of Physics and Astronomy, Tel Aviv University, Tel Aviv 69978, Israel}}
\newcommand{\Edinburgh}{\affiliation{Institute for Astronomy, University of Edinburgh, Royal Observatory, Blackford Hill EH9 3HJ, UK}}
\newcommand{\Birmingham}{\affiliation{Birmingham Institute for Gravitational Wave Astronomy and School of Physics and Astronomy, University of Birmingham, Birmingham B15 2TT, UK}}
\newcommand{\Bath}{\affiliation{Department of Physics, University of Bath, Claverton Down, Bath BA2 7AY, UK}}
\newcommand{\CTIO}{\affiliation{Cerro Tololo Inter-American Observatory, National Optical Astronomy Observatory, Casilla 603, La Serena, Chile}}
\newcommand{\Potsdam}{\affiliation{Institut f\"ur Physik und Astronomie, Universit\"at Potsdam, Haus 28, Karl-Liebknecht-Str. 24/25, D-14476 Potsdam-Golm, Germany}}
\newcommand{\INPE}{\affiliation{Instituto Nacional de Pesquisas Espaciais, Avenida dos Astronautas 1758, 12227-010, S\~ao Jos\'e dos Campos -- SP, Brazil}}
\newcommand{\UNC}{\affiliation{Department of Physics and Astronomy, University of North Carolina, 120 East Cameron Avenue, Chapel Hill, NC 27599, USA}}
\newcommand{\Ohio}{\affiliation{Astrophysical Institute, Department of Physics and Astronomy, 251B Clippinger Lab, Ohio University, Athens, OH 45701-2942, USA}}
\newcommand{\AAS}{\affiliation{American Astronomical Society, 1667 K~Street NW, Suite 800, Washington, DC 20006-1681, USA}}
\newcommand{\MMT}{\affiliation{MMT and Steward Observatories, University of Arizona, 933 North Cherry Avenue, Tucson, AZ 85721-0065, USA}}
\newcommand{\Geneva}{\affiliation{ISDC, Department of Astronomy, University of Geneva, Chemin d'\'Ecogia, 16 CH-1290 Versoix, Switzerland}}
\newcommand{\IUCAA}{\affiliation{Inter-University Center for Astronomy and Astrophysics, Post Bag 4, Ganeshkhind, Pune, Maharashtra 411007, India}}
\newcommand{\CMU}{\affiliation{Department of Physics, Carnegie Mellon University, 5000 Forbes Avenue, Pittsburgh, PA 15213-3815, USA}}
\newcommand{\NAOJ}{\affiliation{Division of Science, National Astronomical Observatory of Japan, 2-21-1 Osawa, Mitaka, Tokyo 181-8588, Japan}}
\newcommand{\IfA}{\affiliation{Institute for Astronomy, University of Hawai`i, 2680 Woodlawn Drive, Honolulu, HI 96822-1839, USA}}
\newcommand{\UCSC}{\affiliation{Department of Astronomy and Astrophysics, University of California, Santa Cruz, CA 95064-1077, USA}}
\newcommand{\Purdue}{\affiliation{Department of Physics and Astronomy, Purdue University, 525 Northwestern Avenue, West Lafayette, IN 47907-2036, USA}}
\newcommand{\Princeton}{\affiliation{Department of Astrophysical Sciences, Princeton University, 4 Ivy Lane, Princeton, NJ 08540-7219, USA}}
\newcommand{\Moore}{\affiliation{Gordon and Betty Moore Foundation, 1661 Page Mill Road, Palo Alto, CA 94304-1209, USA}}
\newcommand{\Durham}{\affiliation{Department of Physics, Durham University, South Road, Durham, DH1 3LE, UK}}
\newcommand{\JHU}{\affiliation{Department of Physics and Astronomy, The Johns Hopkins University, 3400 North Charles Street, Baltimore, MD 21218, USA}}
\newcommand{\Toronto}{\affiliation{David A.\ Dunlap Department of Astronomy and Astrophysics, University of Toronto,\\ 50 St.\ George Street, Toronto, Ontario, M5S 3H4 Canada}}
\newcommand{\Duke}{\affiliation{Department of Physics, Duke University, Campus Box 90305, Durham, NC 27708, USA}}
\newcommand{\NCU}{\affiliation{Graduate Institute of Astronomy, National Central University, 300 Jhongda Road, 32001 Jhongli, Taiwan}}
\newcommand{\Columbia}{\affiliation{Department of Physics and Columbia Astrophysics Laboratory, Columbia University, Pupin Hall, New York, NY 10027, USA}}
\newcommand{\Flatiron}{\affiliation{Center for Computational Astrophysics, Flatiron Institute, 162 5th Avenue, New York, NY 10010-5902, USA}}
\newcommand{\CIERA}{\affiliation{Center for Interdisciplinary Exploration and Research in Astrophysics and Department of Physics and Astronomy, \\Northwestern University, 1800 Sherman Avenue, 8th Floor, Evanston, IL 60201, USA}}
\newcommand{\GeminiNorth}{\affiliation{Gemini Observatory, 670 North A`ohoku Place, Hilo, HI 96720-2700, USA}}
\newcommand{\Keck}{\affiliation{W.~M.~Keck Observatory, 65-1120 M\=amalahoa Highway, Kamuela, HI 96743-8431, USA}}
\newcommand{\UW}{\affiliation{Department of Astronomy, University of Washington, 3910 15th Avenue NE, Seattle, WA 98195-0002, USA}}
\newcommand{\catalyst}{\altaffiliation{LSSTC Catalyst Fellow}}
\newcommand{\USask}{\affiliation{Department of Physics \& Engineering Physics, University of Saskatchewan, 116 Science Place, Saskatoon, SK S7N 5E2, Canada}}
\newcommand{\Thacher}{\affiliation{Thacher School, 5025 Thacher Road, Ojai, CA 93023-8304, USA}}
\newcommand{\Rutgers}{\affiliation{Department of Physics and Astronomy, Rutgers, the State University of New Jersey,\\136 Frelinghuysen Road, Piscataway, NJ 08854-8019, USA}}
\newcommand{\FSU}{\affiliation{Department of Physics, Florida State University, 77 Chieftan Way, Tallahassee, FL 32306-4350, USA}}
\newcommand{\Melbourne}{\affiliation{School of Physics, The University of Melbourne, Parkville, VIC 3010, Australia}}
\newcommand{\ASTROthreeD}{\affiliation{ARC Centre of Excellence for All Sky Astrophysics in 3 Dimensions (ASTRO 3D)}}
\newcommand{\Stromlo}{\affiliation{Mt.\ Stromlo Observatory, The Research School of Astronomy and Astrophysics, Australian National University, ACT 2601, Australia}}
\newcommand{\NCPAS}{\affiliation{National Centre for the Public Awareness of Science, Australian National University, Canberra, ACT 2611, Australia}}
\newcommand{\TAMU}{\affiliation{Department of Physics and Astronomy, Texas A\&M University, 4242 TAMU, College Station, TX 77843, USA}}
\newcommand{\Mitchell}{\affiliation{George P.\ and Cynthia Woods Mitchell Institute for Fundamental Physics \& Astronomy, College Station, TX 77843, USA}}
\newcommand{\ESO}{\affiliation{European Southern Observatory, Alonso de C\'ordova 3107, Casilla 19, Santiago, Chile}}
\newcommand{\ICE}{\affiliation{Institute of Space Sciences (ICE, CSIC), Campus UAB, Carrer
de Can Magrans, s/n, E-08193 Barcelona, Spain}}
\newcommand{\IEEC}{\affiliation{Institut d'Estudis Espacials de Catalunya (IEEC), Edifici RDIT, Campus UPC, 08860 Castelldefels (Barcelona), Spain}}
\newcommand{\Warwick}{\affiliation{Department of Physics, University of Warwick, Gibbet Hill Road, Coventry CV4 7AL, UK}}
\newcommand{\Macquarie}{\affiliation{School of Mathematical and Physical Sciences, Macquarie University, NSW 2109, Australia}}
\newcommand{\AAARC}{\affiliation{Astronomy, Astrophysics and Astrophotonics Research Centre, Macquarie University, Sydney, NSW 2109, Australia}}
\newcommand{\Capodimonte}{\affiliation{INAF - Capodimonte Astronomical Observatory, Salita Moiariello 16, I-80131 Napoli, Italy}}
\newcommand{\INFNNapoli}{\affiliation{INFN - Napoli, Strada Comunale Cinthia, I-80126 Napoli, Italy}}
\newcommand{\ICRANet}{\affiliation{ICRANet, Piazza della Repubblica 10, I-65122 Pescara, Italy}}
\newcommand{\MSU}{\affiliation{Center for Data Intensive and Time Domain Astronomy, Department of Physics and Astronomy,\\Michigan State University, East Lansing, MI 48824, USA}}
\newcommand{\SETI}{\affiliation{SETI Institute,
339 Bernardo Ave, Suite 200, Mountain View, CA 94043, USA}}
\newcommand{\IAIFI}{\affiliation{The NSF AI Institute for Artificial Intelligence and Fundamental Interactions}}
\newcommand{\ANUC}{\affiliation{Department of Astronomy, AlbaNova University Center, Stockholm University, SE-10691 Stockholm, Sweden}}

\newcommand{\Konkoly}{\affiliation{Konkoly Observatory,  CSFK, Konkoly-Thege M. \'ut 15-17, Budapest, 1121, Hungary}}
\newcommand{\ELTE}{\affiliation{ELTE E\"otv\"os Lor\'and University, Institute of Physics, P\'azm\'any P\'eter s\'et\'any 1/A, Budapest, 1117 Hungary}}
\newcommand{\SZTE}{\affiliation{Department of Experimental Physics, University of Szeged, D\'om t\'er 9, Szeged, 6720, Hungary}}
\newcommand{\IdAlta}{\affiliation{Instituto de Alta Investigaci\'on, Sede Esmeralda, Universidad de Tarapac\'a, Av. Luis Emilio Recabarren 2477, Iquique, Chile}}
\newcommand{\Kavli}{\affiliation{Kavli Institute for Cosmological Physics, University of Chicago, Chicago, IL 60637, USA}}
\newcommand{\UofChicago}{\affiliation{Department of Astronomy and Astrophysics, University of Chicago, Chicago, IL 60637, USA}}
\newcommand{\Fermi}{\affiliation{Fermi National Accelerator Laboratory, P.O.\ Box 500, Batavia, IL 60510, USA}}
\newcommand{\Dartmouth}{\affiliation{Department of Physics and Astronomy, Dartmouth College, Hanover, NH 03755, USA}}
\newcommand{\Surrey}{\affiliation{Department of Physics, University of Surrey, Guildford GU2 7XH, UK}}
\newcommand{\NU}{\affiliation{Center for Interdisciplinary Exploration and Research in Astrophysics (CIERA) and Department of Physics and Astronomy, Northwestern University, Evanston, IL 60208, USA}}
\newcommand{\itagaki}{\affiliation{Itagaki Astronomical Observatory, Yamagata 990-2492, Japan}}
\newcommand{\UdChile}{\affiliation{Departamento de Astronomia, Universidad de Chile, Camino El Observatorio 1515, Las Condes, Santiago, Chile}}
\newcommand{\UVA}{\affiliation{Department of Astronomy, University of Virginia, Charlottesville, VA 22904, USA}}
\author[0000-0002-4022-1874]{Manisha Shrestha}
\UA

\author[0000-0002-4924-444X]{K. Azalee Bostroem}
\catalyst\UA

\author[0000-0003-4102-380X]{David J. Sand}
\UA

\author[0000-0002-0832-2974]{Griffin Hosseinzadeh}
\UA

\author[0000-0003-0123-0062]{Jennifer E. Andrews}
\GeminiNorth

\author[0000-0002-7937-6371]{Yize Dong \begin{CJK*}{UTF8}{gbsn}(董一泽)\end{CJK*}}
\UCD

\author[0000-0003-2744-4755]{Emily Hoang}
\UCD

\author[0000-0003-0549-3281]{Daryl Janzen}
\USask

\author[0000-0002-0744-0047]{Jeniveve Pearson}
\UA

\author[0000-0001-5754-4007]{Jacob E. Jencson}
\IPAC

\author[0000-0001-9589-3793]{M.~J. Lundquist}

\Keck
\author[0009-0008-9693-4348]{Darshana Mehta}
\UCD
\author[0000-0002-7352-7845]{Aravind P.\ Ravi}
\UCD

\author[0000-0002-7015-3446]{Nicol\'as Meza Retamal}
\UCD

\author[0000-0001-8818-0795]{Stefano Valenti}
\UCD

\author[0000-0001-6272-5507]{Peter J. Brown}
\TAMU

\author[0000-0001-8738-6011]{Saurabh W.\ Jha}
\Rutgers

\author[0000-0002-9209-2787]{Colin Macrie}
\Rutgers
\author[0000-0002-9454-1742]{Brian Hsu}
\UA

\author[0000-0003-4914-5625]{Joseph Farah}
\LCO 
\UCSB

\author[0000-0003-4253-656X]{D.\ Andrew Howell}
\LCO\UCSB
\author[0000-0001-5807-7893]{Curtis McCully}
\LCO

\author[0000-0001-9570-0584]{Megan Newsome}
\LCO 
\UCSB

\author[0000-0003-0209-9246]{Estefania Padilla Gonzalez}
\LCO
\UCSB
\author[0000-0002-7472-1279]{Craig Pellegrino}
\UVA
\author[0000-0003-0794-5982]{Giacomo Terreran}
\LCO 
\UCSB

\author[0000-0003-3108-1328]{Lindsey Kwok}
\Rutgers

\author[0000-0001-5510-2424]{Nathan Smith}
\UA

\author[0009-0002-5096-1689]{Michaela Schwab}
\Rutgers
\author{Aidan Martas}
\UCD

\author[0000-0002-0810-5558]{Ricardo R. Munoz}
\UdChile

\author[0000-0003-0105-9576]{Gustavo E. Medina}
\Toronto

\author[0000-0002-9110-6163]{Ting S. Li}
\Toronto

\author{Paula Diaz}
\UdChile

\author[0000-0002-1125-9187]{Daichi Hiramatsu}
\CfA
\IAIFI

\author{Brad E. Tucker}
\Stromlo

\author[0000-0003-1349-6538]{J. C. Wheeler}
\UT

\author[0000-0002-7334-2357]{Xiaofeng Wang}
\Tsinghua

\author{Qian Zhai}
\YNAO

\author[0000-0002-8296-2590]{Jujia Zhang}
\YNAO
\ICEY

\author[0000-0002-3884-5637]{Anjasha Gangopadhyay}
\OKC

\author[0000-0002-6535-8500]{Yi Yang}
\Tsinghua

\author[0000-0003-2375-2064]{Claudia P. Guti\'{e}rrez}
\IEEC\ICE

\begin{abstract}

We present high-cadence photometric and spectroscopic observations of supernova (SN) 2024ggi, a Type II SN with flash spectroscopy features, which exploded in the nearby galaxy NGC~3621 at $\sim$7 Mpc. 
The light-curve evolution over the first 30 hours can be fit by two power law indices with a break after 22 hours, rising from $M_V \approx -12.95$ mag at +0.66 days to $M_V \approx -17.91$ mag after 7 days. In addition, the densely sampled color curve shows a strong blueward evolution over the first few days and then behaves as a normal SN II with a redward evolution as the ejecta cool.
Such deviations could be due to interaction with circumstellar material (CSM).
Early high- and low-resolution spectra clearly show high-ionization flash features from the first spectrum to +3.42 days after the explosion. From the high-resolution spectra, we calculate the CSM velocity to be 37 $\pm~4~\mathrm{km\,s^{-1}} $. We also see the line strength evolve rapidly from 1.22 to 1.49 days in the earliest high-resolution spectra. Comparison of the low-resolution spectra with CMFGEN models suggests that the pre-explosion mass-loss rate of SN\,2024ggi falls in a range of $10^{-3}$ to $10^{-2}$ M$_{\sun}$ yr$^{-1}$, which is similar to that derived for SN\,2023ixf. However, the rapid temporal evolution of the narrow lines in the spectra of SN~2024ggi ($R_\mathrm{CSM} \sim 2.7 \times 10^{14} \mathrm{cm}$) could indicate a smaller spatial extent of the CSM than in SN~2023ixf ($R_\mathrm{CSM} \sim 5.4 \times 10^{14} \mathrm{cm}$) which in turn implies lower total CSM mass for SN~2024ggi.

\end{abstract}



\section{Introduction} \label{sec:intro}

Nearby core-collapse supernovae (SNe) 
offer an unparalleled opportunity to learn about their progenitor star systems, pre-explosion environments, and the explosions themselves. The proliferation of high-cadence time-domain surveys tied to rapid-response facilities has yielded exquisite data sets, leading to sizeable advances and new questions, many of them centered around the circumstellar material (CSM) associated with the star at explosion and the effects this material may have on the early light curves and spectra.

While it is well established that red supergiant (RSG) stars are the progenitors of normal Type II SNe (SNe II) \citep[][and references therein]{Smartt_2009, Smartt_2015}, the final stages of RSG evolution are still uncertain. Shock breakout (SBO) is expected when the optical depth drops below $\sim c/v_{sh}$ where $c$ is the speed of light and $v_{sh}$ is the shock velocity \citep{Waxman_2017_SBO_theory}. This breakout produces flashes of X-ray/UV on short timescales (seconds to hours) and UV/optical emission on $\sim$day timescales from the expanding cooling envelope. The SBO duration could be attributed to light-travel time across the stellar surface which can provide a constraint on the stellar radius \citep{Goldberg_2022_SBO, Waxman_2017_SBO_theory}. However, in the presence of dense CSM these time scales can increase to days \citep[e.g][]{Chugai_2004_sbo,smith_2007_sbo,Ofek_2010_SBO,Dessart_2017_sbo,Haynie_2021_sbo}. This SBO from dense CSM can produce ionized narrow emission lines in their spectra during this time.

A significant fraction of SNe II display narrow flash features in the days after explosion \citep[e.g.,][]{GalYam_2014,Smith_2015,Khazov_2016,Yaron_2017, Hosseinzadeh_2018,2020MNRAS.498...84Z,Tartaglia_2021,Terreran_2022, Bostroem_2023_23ixf,Jacobson-Galan_2023,Bruch_2023,Andrews_2024}, which signify interaction with the material from pre-supernova mass loss of the progenitor star. Interestingly, this material seems to be confined ($\lesssim$10$^{15}$ cm) and the estimated mass-loss rates from these early SN observations indicate $\dot{M} = 10^{-3}-10^{-2}\ M_\sun\ \mathrm{yr}^{-1}$ \citep[][for a recent compilation]{JG24}, orders of magnitude higher than values inferred for quiescent mass loss in RSGs \citep[e.g.][]{deJager_1988,Ekstrom_2012,Beasor_2020}.  Such high mass loss in the final years before explosion can also affect the early light-curve evolution, causing it to deviate from standard shock-breakout scenarios \citep[e.g. see recent results on SN~2023ixf;][]{Hosseinzadeh_2023_23ixf,Hiramatsu2023ApJ...955L...8H, Li_2024_23ixf,Zimmerman_2024_2023ixf}.

\begin{figure*}
    \centering
    \includegraphics[scale=0.15]{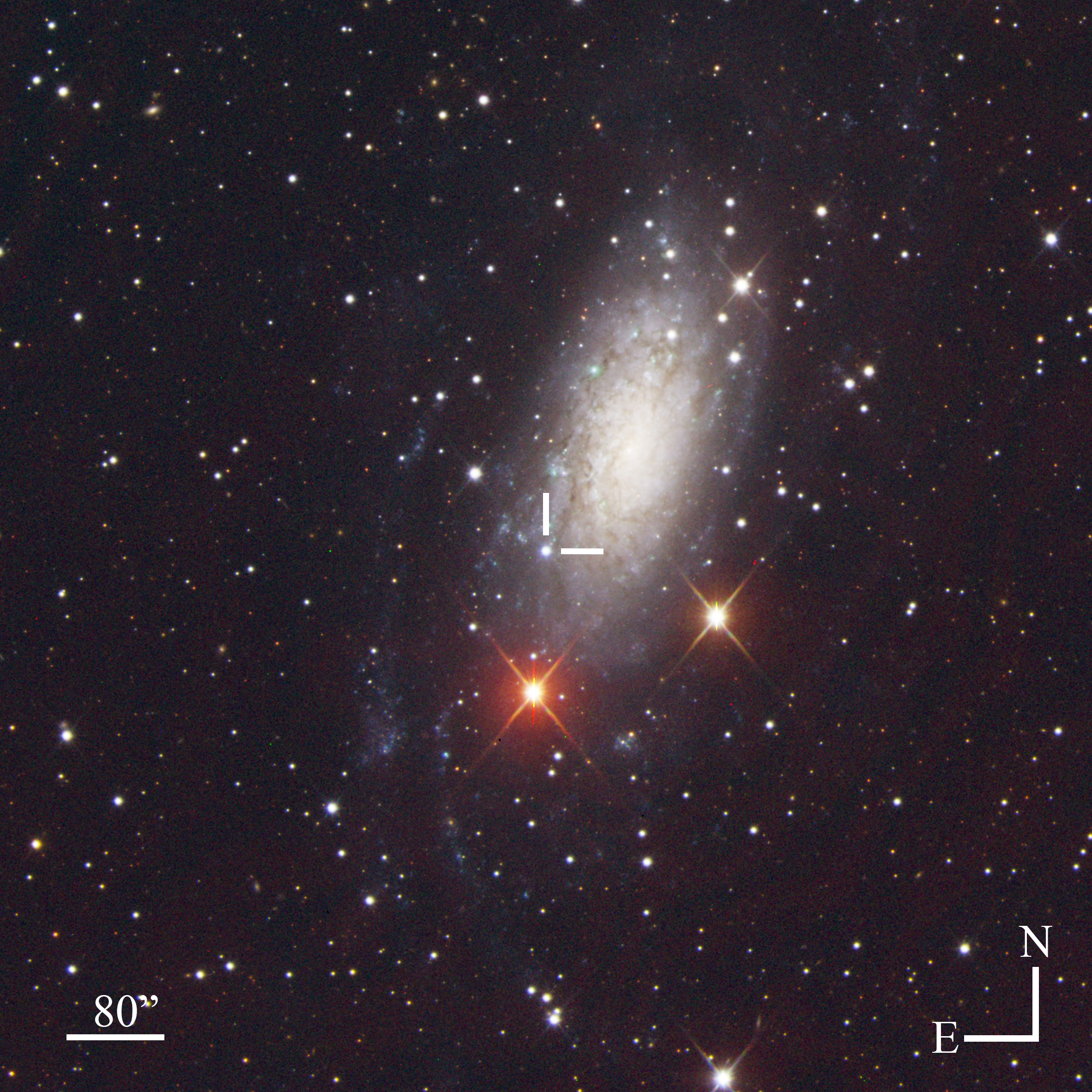}
    \caption{Composite gri image of SN~2024ggi and its host galaxy NGC~3621 taken by Las Cumbres Observatory on MJD 60412.46 (+1.7 days after the explosion).}
    \label{fig:hostgalaxy}
\end{figure*}

Here, we present a new study of a very nearby core-collapse SN that showed prominent signs of interaction with dense CSM.  SN~2024ggi was discovered by the Asteroid Terrestrial-Impact Last Alert System \citep[ATLAS][]{Tonry_2011,Tonry_2018,Smith_2020,Srivastav_2024,Chen_2024_24ggi} on 2024-04-11 03:22:36 UT \citep[MJD 60411.14;][]{Tonry_2024} in NGC 3621 (see Figure~\ref{fig:hostgalaxy}). It was classified as a SN II with prominent flash features on 2024-04-11 14:35:47 UT \citep[MJD 60411.61;][]{2024_Zhai}. A tight non-detection limit was obtained by the Gravitational-wave Optical Transient Observer \citep[GOTO;][]{2022_Steeghs} on 2024-04-10 10:56:25 UT (MJD 60410.46), with a limiting magnitude of $L=19.5$ mag \citep{2024_Killestein}, less than one day before the first ATLAS detection. We use the midpoint of the last non-detection (MJD 60410.46), and first detection (MJD 60411.14) as our explosion epoch throughout this work (i.e., MJD 60410.80 $\pm$ 0.34) and all the phases in this paper are calculated based on this value. We also adopt the Cepheid-based distance to NGC3621 of $D=7.24\pm0.20$ Mpc \citep{Saha06}, identical to that used in \citet{Wynn_2024_ggi}.  Several basic parameters associated with SN~2024ggi are shown in Table~\ref{tab:results}.

A few other studies have presented results on SN~2024ggi.  In \citet{Wynn_2024_ggi}, the early light curve and low-dispersion flash spectroscopy were presented and placed in context with the larger population of CSM-interacting SNe II.  Two high-resolution spectra taken the night after discovery were presented in \citet{Pessi_2024}, showing evolution in the flash features on $\sim$7 hour time scales. An early multi-band light curve was presented in \citet{Chen_2024_24ggi}, while an analysis of the progenitor RSG was presented in \citet{Xiang_2024_24ggi}. Analysis of the archival Hubble Space Telescope (HST) and \textit{Spitzer} Space Telescope images obtained over 20 years before the SN exploded indicates that the progenitor of SN\,2024ggi is consistent with a red supergiant (RSG) star with a temperature of $T = 3290^{+19}_{-27}$ K, radius of $R = 887^{+60}_{-51} R_\sun$, and initial mass of $M = 13 \pm 1 M_\sun$ \citep{Xiang_2024_24ggi}. \citet{Zhang_2024_24ggi} presented high cadence early time spectra with flash features. From their analysis of these spectra they find the CSM extends to $4 \times 10^{14}$ cm with a mass loss rate of $5 \times 10^{-3} M_\sun {\rm yr}^{-1}$ and velocity of unshocked CSM to be between 20 and 40 km s$^{-1}$. \citet{Soker_2024_24ggi} compared the observational data and theoretical work on SN~2024ggi with various explosion mechanisms and found that the jittering jets explosion mechanism is favored for this SN explosion.

This paper presents an extremely high-cadence early light curve and spectroscopic time sequence of SN~2024ggi, including a series of high-resolution spectra (Section~\ref{sec:reduce-obs}).  After estimating the line-of-sight extinction based on the measurements of the pseudo-equivalent width of the Na{\sc ID} lines (Section~\ref{sec:extinct}), we derive and interpret the intrinsic photometric and color evolution of SN\,2024ggi in Section~\ref{sec:LCsection},
including a search for precursor outbursts, the remarkable evolution seen on the first day, and a shock-cooling analysis.  We analyze the spectroscopic sequence of SN~2024ggi in Section~\ref{sec:spec_analysis}, with an emphasis on the flash features, their evolution, and the implications for the CSM surrounding the progenitor at the time of explosion.  In Section~\ref{sec:SN23ixf}, we compare SN~2024ggi with the recent SN II in M101, SN~2023ixf, which also displayed well-observed flash features, is located at nearly the same distance, and has a similar-quality data set.  We summarize and conclude the results in Section~\ref{sec:conclude}.



\section{Observations \& Data reduction} \label{sec:reduce-obs}
\begin{table}
 \caption{Properties of SN~2024ggi} \label{tab:results}
 \begin{tabular}{ll}
    \hline
    Parameter & Value \\
    \hline
    R.A. (J2000) & 11:18:22.09 \\
    Dec. (J2000) & $-$32:50:15.3 \\
    Last Nondetection (MJD) & 60410.46 \\ 
    First Detection (MJD) & 60411.14 \\
    Explosion Epoch (MJD)\tablenotemark{a} & 60410.80 \\
    Redshift ($z$)\tablenotemark{b} & 0.00221  \\
    Distance\tablenotemark{c} & 7.24  $\pm$ 0.2 Mpc\\
    Distance modulus ($\mu$)\tablenotemark{c} & 29.30 mag\\
    $E(B-V)_\mathrm{MW}$ \tablenotemark{a}& $0.054 \pm 0.020$ mag\\
    $E(B-V)_\mathrm{cloud}\tablenotemark{a}$ & $0.066 \pm 0.020$ mag\\
    $E(B-V)_\mathrm{host}\tablenotemark{a}$ & $0.034 \pm 0.020$ mag\\
    $E(B-V)_\mathrm{tot}$\tablenotemark{a} & $0.154 \pm 0.035$ mag\\
    Peak Magnitude ($V_{\mathrm{max}}$) & $-17.72$ mag\\
    Time of $V_{\mathrm{max}}$ (MJD) & 60417.82 \\
    Rise time ($V$) & 7 days \\
    \hline
 \end{tabular}
 \tablenotetext{a}{mid point of last nondetection and first detection}
 \tablenotetext{b}{from the \ion{Na}{1}~D lines of the host galaxy}
 \tablenotetext{c}{from \citet{Saha06}}
\end{table}

\begin{table*}
 \caption{Log of Spectroscopic Observations}
 \begin{tabular}{ c c c c c c}
    \hline
    Date (UTC) & MJD &  Telescope & Instrument & Phase (d)  & Exp (s) \\
    \hline
    2024-04-11 14:35:47 & 60411.61 & Lijiang 2.4m  & YFOSC & 0.81  & 1800\\
    2024-04-11 23:31:08 & 60411.98 & SALT   & RSS &  1.18 & 2393 \\
    2024-04-11 23:46:57 & 60411.99 & Gemini-S & GMOS-S & 1.19& 300 \\ 
    2024-04-12 00:33:44 & 60412.02 & Magellan-Clay & MIKE &1.22 & 300 \\ 
     2024-04-12 07:10:06 & 60412.29 & Magellan-Clay & MIKE &1.49 & 300 \\
    2024-04-12 12:51:33 & 60412.53 & COJ   & FLOYDS & 1.73  & 1800 \\
    2024-04-12 23:23:34 & 60412.97 & SALT   & RSS &2.17   & 2393 \\
    2024-04-13 05:15:11 & 60413.22 & Bok   & B\&C & 2.42  & 2100 \\
    2024-04-13 23:04:55 & 60413.96 & SALT   & HRS & 3.16  & 1300 \\ 
    2024-04-14 05:26:54 & 60414.22 & Bok   & B\&C & 3.42  & 1400 \\
    2024-04-15 04:59:09 & 60415.21 & Bok   & B\&C & 4.41  & 1400 \\
    2024-04-16 06:46:05 & 60416.28 & LCO Faulkes 2m   & FLOYDS & 5.48  & 600 \\
    2024-04-17 23:13:39  & 60417.97 & SALT   & RSS &7.16   & 1893 \\
    2024-04-19 22:40:23 & 60419.94 & SALT   & HRS & 9.14  & 1300 \\
    2024-04-19 23:03:13 & 60419.96 & SALT   & HRS&  9.16 & 1300 \\
   2024-04-20 06:00:18 & 60420.25 & LCO Faulkes 2m   & FLOYDS & 9.45  & 600 \\
   2024-04-20 22:50:20 & 60420.95 & SALT   & RSS & 10.15  & 1893 \\
   2024-04-21 07:36:22 & 60421.31 & LCO Faulkes 2m   & FLOYDS & 10.51  & 600 \\
   2024-04-23 06:38:39 & 60423.27 & LCO Faulkes 2m   & FLOYDS & 12.47  & 600 \\
    \hline
 \end{tabular}
 
 \label{tab:specInst}
\end{table*}

Immediately after the discovery of SN~2024ggi, we began an intensive, high-cadence photometric and spectroscopic campaign, as described in the following sections.

\subsection{Photometry}
In the early hours after the discovery announcement, we acquired a nearly continuous multiband photometric sequence with the Distance Less Than 40 Mpc (DLT40; \citealt{Tartaglia_2018}) survey's two southern telescopes -- the PROMPT5 0.4~m telescope at the Cerro Tololo International Observatory and the PROMPT-MO 0.4~m telescope at the Meckering Observatory in Australia, operated by the Skynet telescope network \citep{Reichart_2005}. Data were taken in $B$, $V$, $g$, $r$, and $i$, as well as a wide-band configuration. In this wide mode, the PROMPT5 telescope observations are filterless (`Open') and the PROMPT-MO telescope utilizes a broadband `Clear' filter, both of which are calibrated to the Sloan Digital Sky Survey $r$ band \citep[see][for further reduction details]{Tartaglia_2018}. Aperture photometry on the DLT40 multiband $B$, $V$, $g$, $r$, and $i$ images was performed using Photutils \citep{Bradley_2022} and was then calibrated to the American Association of Variable Star Observers (AAVSO) Photometric All-Sky Survey \citep[APASS;][]{Henden_2009}.

Further high-cadence multiband observations were taken with the Las Cumbres Observatory telescope network of 0.4~m and 1.0~m telescopes \citep{Brown_2013} in the $U$, $B$, $V$, $g$, $r$, and $i$ bands from the Global Supernova Project. The images were reduced using a PyRAF-based photometric reduction pipeline \citep{Valenti_2016}.  Apparent magnitudes were calibrated using the APASS catalog for $g$, $r$, and $i$ and using Landolt standard fields observed with the same telescope on the same night for $U$, $B$, and $V$.

We used the ATLAS forced photometry service \citep{Tonry_2018,Smith_2020} to obtain further high-cadence photometry in two filters, cyan ($c$) and orange ($o$), which are roughly equivalent to Pan-STARRS filters $g+r$ and $r+i$, respectively. 

Ultraviolet and optical images were obtained with the Ultraviolet/Optical Telescope \citep[UVOT;][]{Roming05} on the Neil Gehrels \textit{Swift} Observatory \citep{Gehrels_2004} under a guest investigator program (PI: A. P. Ravi) that requested high-cadence (every $\sim$6 hr) observations at early times. Further standard ToO requests to \textit{Swift} at a more moderate cadence were obtained after several days. The UVOT images were reduced using the High-Energy Astrophysics software (HEASoft\footnote{https://heasarc.gsfc.nasa.gov/docs/software/heasoft/}). The source region is centered at the position of the SN with an aperture size of 3$\arcsec$ and the background is measured from a region without contamination from other stars with an aperture size of 5$\arcsec$. Several epochs near the peak were observed with a faster readout setup (UVOT hardware mode: 0x0378) to mitigate saturation. Zero points for photometry were chosen from \cite{Breeveld_2010} with time-dependent sensitivity corrections updated in 2020. 

\begin{figure*}
    \centering
    \includegraphics[width=\textwidth]{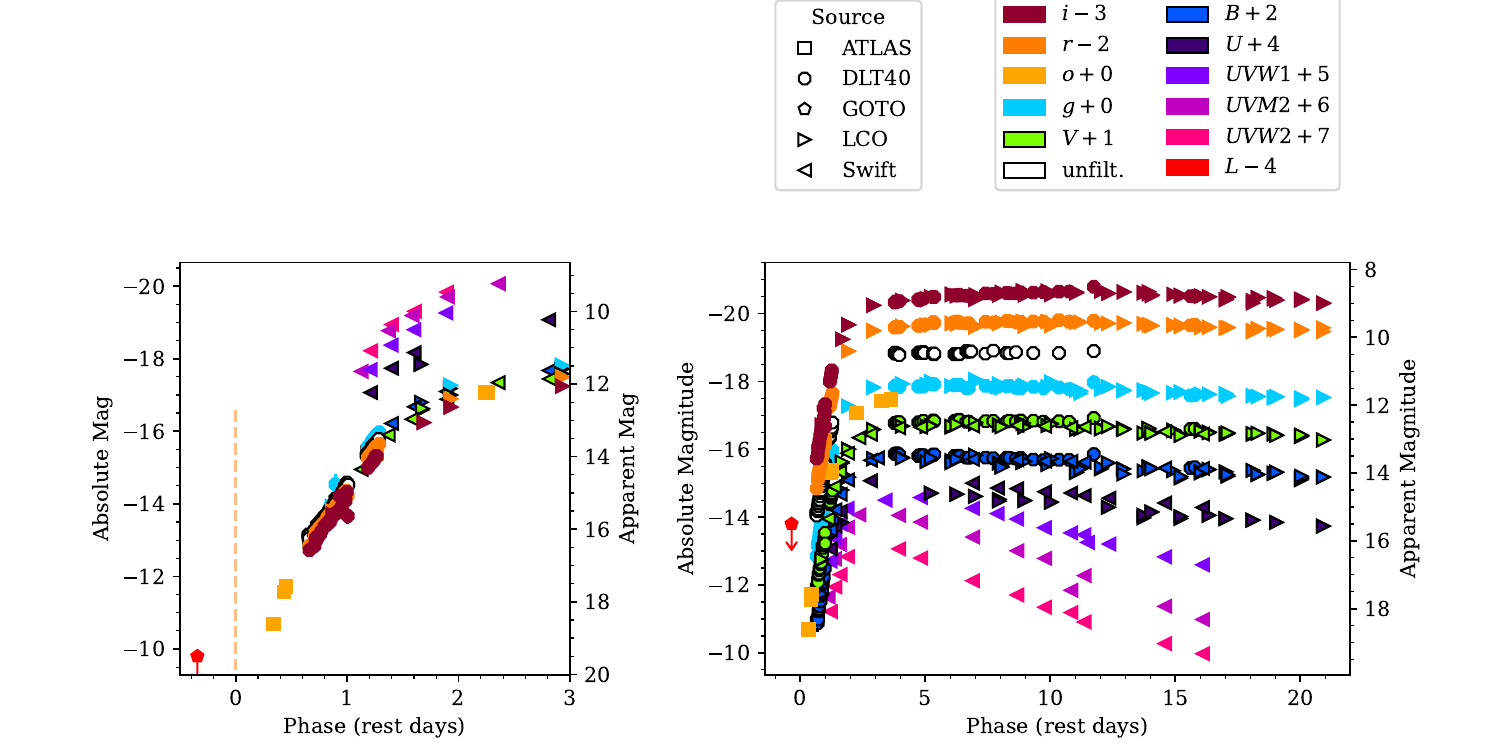}
    \caption{Left: Multiwavelength photometry of SN~2024ggi for the first 2.5 days in absolute and extinction-corrected (Milky Way + host) apparent magnitudes with no filter shifts. The dashed orange line represents the assumed explosion epoch. Right: The full multiwavelength light curve of SN~2024ggi extending to 21 days, including data from ATLAS, DLT40, Swift, Las Cumbres Observatory, and the last non-detection from GOTO \citep{Killestein_2024_goto}. The data is well sampled throughout the first 20 days including the very early rise and plateau. (The data used to create this figure are available in the published article.)}
    \label{fig:lv}
\end{figure*}

The complete early light curve from discovery until the SN is firmly on the plateau (MJD 60431.77) is shown in Figure~\ref{fig:lv}. A thorough analysis of the early photometry can be found in Section~\ref{sec:LCsection}.

\subsection{Spectroscopy}

\begin{figure*}
    \centering
    \includegraphics[width=\textwidth]{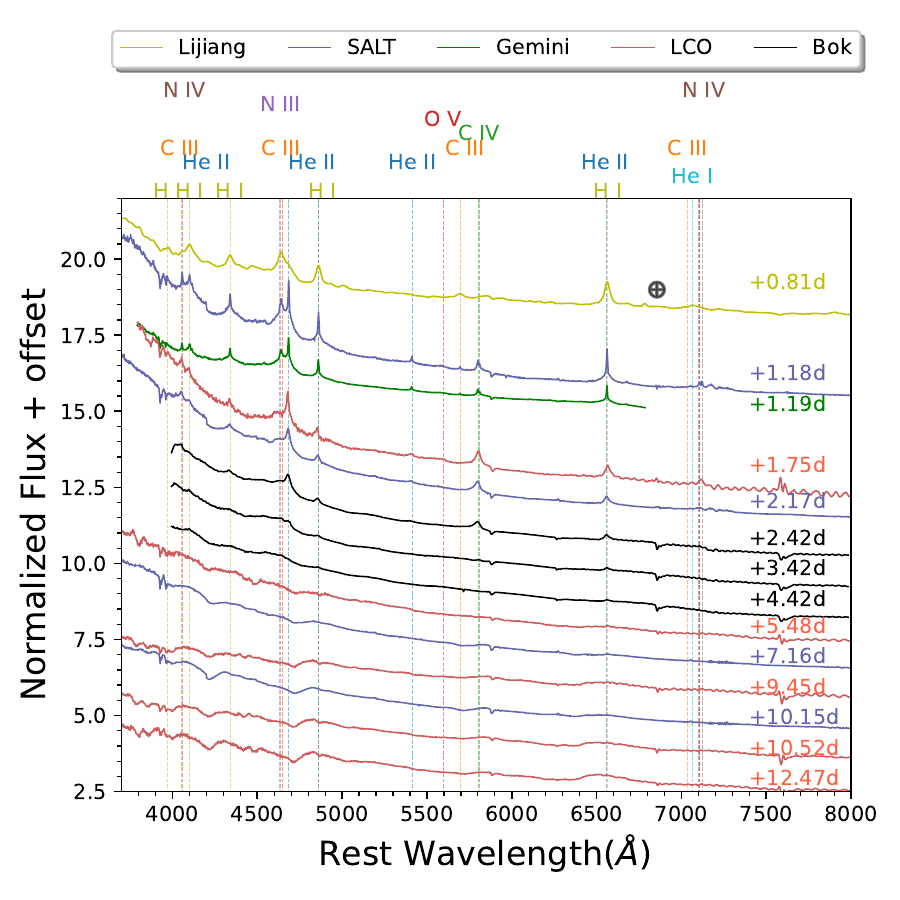}
    \caption{Optical spectral sequence of SN~2024ggi starting from +0.9 days \citep{2024_Zhai} to +12.6 days after the explosion. Spectra from different instruments are indicated by different colors, as indicated in the legend, with the phases indicated on the right. 
    The early spectra show flash features which disappear by roughly +3.5 days. All spectra will be made available on WISeREP (\url{https://www.wiserep.org}). (The data used to create this figure are available in the published article.)  }
    \label{fig:spec-seq}
\end{figure*}

We present a log of spectroscopic observations in Table~\ref{tab:specInst}, and the full low-dispersion spectroscopic sequence is shown in Figure~\ref{fig:spec-seq}. Portions of our higher-resolution spectra are presented later in Section~\ref{sec:highres}. 

The first spectrum of SN~2024ggi was taken with the Yunnan Faint Object Spectrograph and Camera (YFOSC) on the Lijiang 2.4m telescope \citep{yfosc}, which was used to classify it as a young SN II with flash-spectroscopy features \citep{2024_Zhai,Zhang_2024_24ggi}.  We have downloaded the public spectrum from the Transient Name Server to add to our analysis.

Multiple spectra were obtained with the Robert Stobie Spectrograph (RSS) on the Southern African Large Telescope \citep[SALT;][]{SALT}, including one taken hours after the initial YFOSC spectrum exhibiting clear evolution in the flash features. The SALT data were reduced using a custom pipeline based on the PySALT package \citep{Crawford_2010}.

Another early spectrum was obtained with the Gemini Multi-Object Spectrograph (GMOS; \citealp{hook_2004,gimeno_2016}) on the 8.1\,m Gemini South Telescope using the B600 grating. Data were reduced using the Data Reduction for Astronomy from Gemini Observatory North and South ({\tt DRAGONS}) package \citep{Labrie_2019}, using the recipe for GMOS long-slit reductions. This includes bias correction, flat-fielding, wavelength calibration, and flux calibration. 

Further spectra were obtained with the FLOYDS spectrographs \citep{Brown_2013} on the Las Cumbres Observatory's 2m Faulkes Telescopes North and South (FTN/FTS) as part of the Global Supernova Project. One-dimensional spectra were extracted, reduced, and calibrated following standard procedures using the FLOYDS pipeline \citep{Valenti_2014}.

We also include spectra taken with the Boller and Chivens (B\&C) Spectrograph on the University of Arizona's Bok 2.3~m telescope located at Kitt Peak Observatory, which were reduced in a standard way with IRAF \citep{iraf1,iraf2} routines.

Two sets of moderate- to high-resolution spectra were also taken.
First, two epochs of echelle spectra were taken in a single night separated by 7 hours with the Magellan Inamori Kyocera Echelle (MIKE) double echelle spectrograph at the Magellan Clay Telescope
at Las Campanas Observatory in Chile \citep{Bernstein_2003} with resolution of $R \approx 22{,}600$ for the red side (4900-9500 $\AA$), and $R \approx 28{,}000$ for the blue side (3350-5000 $\AA$). \citet{Pessi_2024} have also presented an analysis of these two spectra. The spectra were reduced using the latest version of the MIKE
pipeline using CarPy \citep{Kelson_2000,Kelson_2003}. A barycentric velocity correction was applied to the data.

Another set of echelle spectra were taken with the SALT High Resolution Spectrograph \citep[HRS;][]{SALT_HRS} in moderate resolution mode ($R \approx 40{,}000$).  These data were reduced with the standard MIDAS HRS pipeline\footnote{\url{https://astronomers.salt.ac.za/software/hrs-pipeline/}}, with steps including flat-fielding, wavelength calibration, and extraction \citep{Kniazev16,Kniazev17}. A barycentric velocity correction was applied to the data.  We found that the \ion{Na}{1}~D lines from the HRS data are systematically shifted by $-4\ \mathrm{km\ s^{-1}}$ compared to the MIKE data. As this is less than a resolution element, for consistency, we shift all the HRS data to the MIKE data for consistency.

\section{Line of Sight Extinction}\label{sec:extinct}

\begin{figure}
    \centering
    \includegraphics[width=\columnwidth]{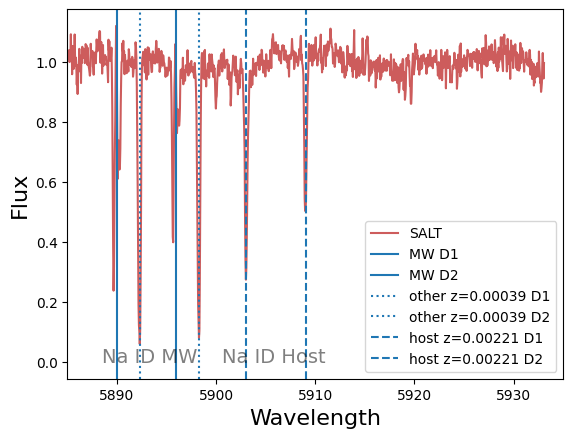}
    \caption{High-resolution data from SALT HRS observed on MJD 60413.96 (+3.16 days) focused on the \ion{Na}{1}~D lines of the host galaxy and Milky Way. There are six different absorption features corresponding to \ion{Na}{1}~D1 and D2 lines for the Milky Way, the host galaxy, and an intervening cloud, shown by blue lines for different redshifts. The total equivalent width of these lines measured from all the high-resolution spectra was used to calculate the extinction towards SN~2024ggi.}
    \label{fig:extinction}
\end{figure}

We clearly detect \ion{Na}{1}~D absorption lines in our high-resolution spectra, which are known to correlate with interstellar dust extinction \citep{Richmond_1994,Munari_1997,Poznanski_2012}. As shown in Figure~\ref{fig:extinction}, there are components associated with Milky Way absorption for both \ion{Na}{1}~D1 and D2 lines. The Milky Way absorption is composed of at least two component clouds, but we treat them as a single feature while calculating the equivalent width. In addition, there is another feature at $z=0.00039$, which we interpret as another intervening cloud, as well as absorption features associated with the host galaxy. We calculated the equivalent width of \ion{Na}{1}~D1 and \ion{Na}{1}~D2 absorption features for the Milky Way, the intervening dust cloud, and the host galaxy using the \texttt{splot} task from IRAF for all five high-resolution spectra from MIKE and HRS. The equivalent width was used in Equation 9 of \citet{Poznanski_2012} to calculate the color excess of each component, and a renormalization factor of 0.86 was applied following \citet{Schlafly_2011}. From this calculation, we find $E(B-V)_{\rm MW} = 0.054 \pm 0.020$, $E(B-V)_{\rm cloud} = 0.066 \pm 0.020$, and $E(B-V)_{\rm host} = 0.034 \pm 0.020$. The uncertainty in the derived extinction is dominated by the systematic uncertainty estimated by \citet{Poznanski_2012}. Note that the Milky Way value from \citet{Schlafly_2011} is $E(B-V)_{\rm MW} = 0.0698$ mag, consistent with our \ion{Na}{1}~D-based measurement within the uncertainties. Thus the total extinction for SN~2024ggi is $E(B-V)_{\rm tot} = 0.154 \pm 0.035$ mag, which we adopt for this paper. This value is consistent with values calculated by \citet{Pessi_2024} and \citet{Wynn_2024_ggi}.

\section{Light curve \& Color Evolution} \label{sec:LCsection}

The high cadence light-curve evolution of SN~2024ggi obtained by ATLAS, Swift, Las Cumbres, and DLT40 is presented in Figure~\ref{fig:lv} with the right panel showing all data up to +21 days with respect to our assumed explosion epoch. In the left panel, we focus on the light curve over the first 2.5 days. The peak absolute magnitude in the $V$ band is  $-17.72$ mag, 7 days after explosion on MJD 60417.8.  In this section, we search for precursor emission at the SN position and explore the early light-curve and color evolution ($\lesssim$1--2 d) with our extremely high-cadence data.

\subsection{Precursor Search}

It is clear that there is dense, confined CSM around many of the progenitors of SNe II \citep{Forster2018NatAs...2..808F,Khazov_2016,Morozova_2018, Bruch_2021, Bruch_2023}, indicating intense mass loss from the progenitor in the months to years before the explosion. This mass loss may produce observable precursor outburst activity. Precursors are commonly observed in  SNe IIn \citep{Ofek14,Strotjohann21}, which are characterized by narrow hydrogen emission lines. In contrast, precursor emission has only been identified in one normal SN II \citep[2020tlf;][]{Jacobson-Galan2022}. Other efforts to search for precursors in SNe II have yielded only nondetections \citep[e.g.,][]{Johnson2018MNRAS.480.1696J}, including for the nearby SN~2023ixf, whose pre-SN site had been observed for many years by several major surveys \citep{Neustadt2024MNRAS.527.5366N,Hiramatsu2023ApJ...955L...8H, Dong2023ApJ...957...28D, Ransome2024ApJ...965...93R,Rest24}.

\begin{figure}
    \centering
    \includegraphics[width=\columnwidth]{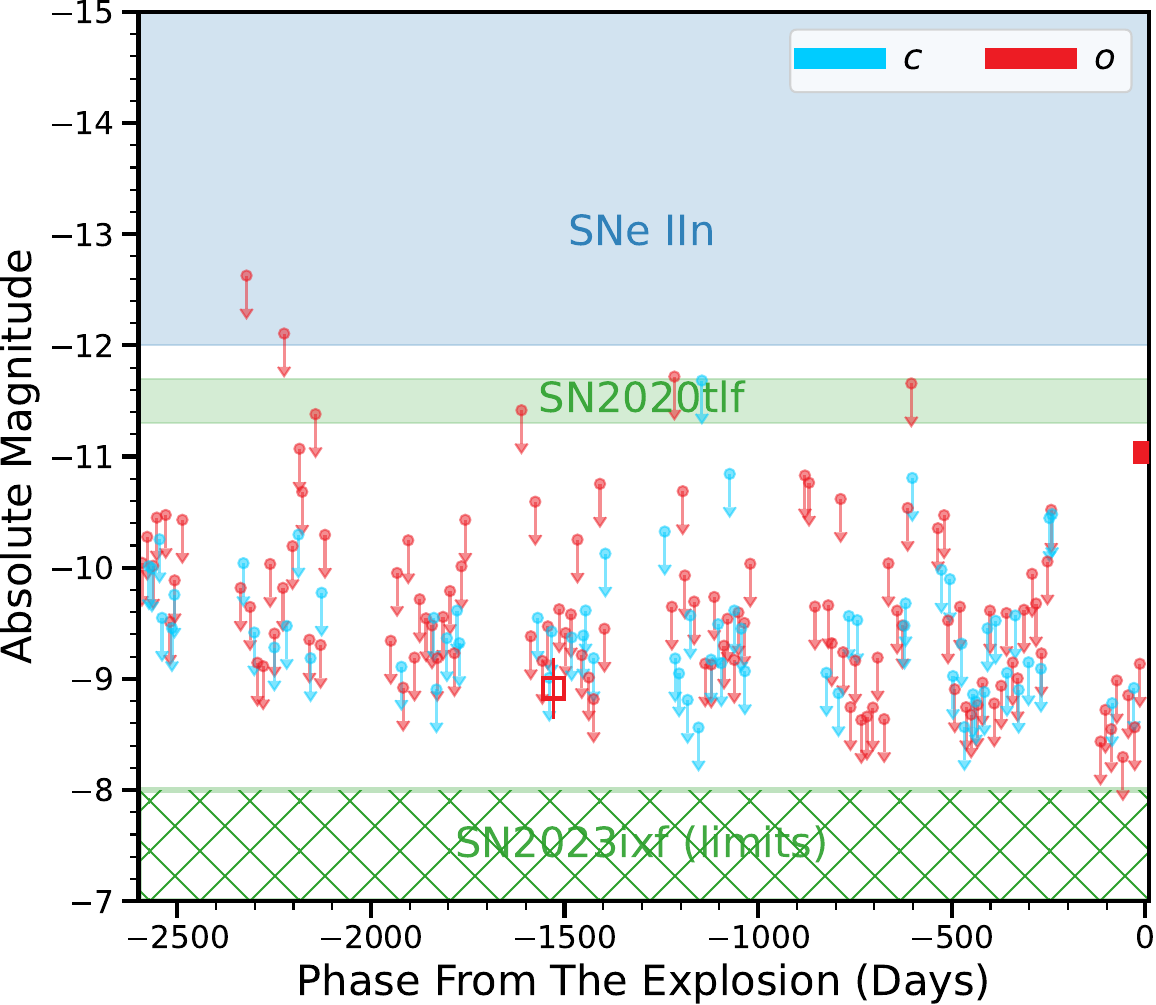}
    \caption{Nondetection limits on the pre-explosion activity of SN~2024ggi. The typical brightness of pre-explosion outbursts of SNe IIn is indicated by the blue area \citep{Ofek14,Strotjohann21}. The brightness of precursor emission observed in Type II SN~2020tlf is marked with the green area \citep{Jacobson-Galan2022}, while the nondetection limit on the pre-explosion activity of Type II SN~2023ixf is represented by the green hatched area \citep{Dong2023ApJ...957...28D,Ransome2024ApJ...965...93R,Rest24}.}
    \label{fig:precursor}
\end{figure}

The position of SN~2024ggi has been monitored by ATLAS since 2017. We obtained forced photometry at the SN site from the ATLAS forced photometry server \citep{Shingles2021}, which was then stacked in windows of 15 days to achieve deeper limits using the method presented in \cite{Young2022}. We adopted a signal-to-noise threshold of 3 for source detections and a signal-to-noise threshold of 5 for calculating the upper limits, as recommended by \cite{Masci2011ComputingFU}. The results are presented in Figure \ref{fig:precursor}. 
Therefore, we conclude that precursor emission is not detected for SN~2024ggi down to $\sim$$-9$ mag. 

If there were precursor activities for SN~2024ggi, they would have to be faint and only last for a short duration. 
Although the nondetection of precursors for SN~2024ggi is not as deep as that for SN~2023ixf, it is still $\sim$2 mag deeper than the precursor detected in SN~2020tlf. 
The diversity of precursor activities observed in these SNe suggests that multiple physical mechanisms are responsible for the dense and confined CSM around the progenitors of SNe II, as suggested by \citet{Dong2023ApJ...957...28D}.

\subsection{Temperature \& Radius}
\begin{figure}
    \centering
    \includegraphics[width=\columnwidth]{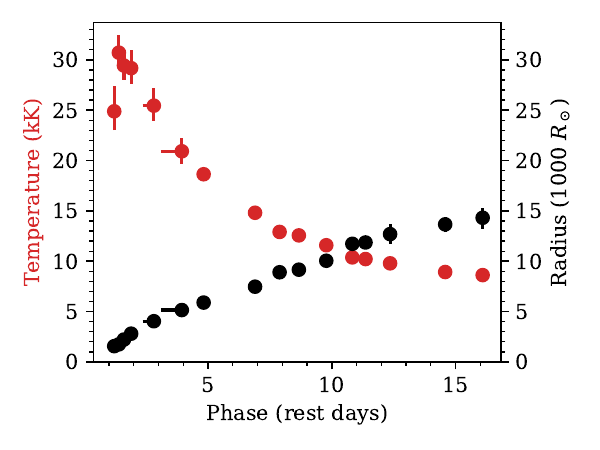}
    \caption{Temperature (red) and photospheric radius (black) evolution of SN~2024ggi derived from fitting a blackbody SED to the multiband photometry.} 
    \label{fig:rad-temp}
\end{figure}

We construct a bolometric light curve using the multiband photometry obtained for SN~2024ggi. During these earliest phases, the spectral energy distribution (SED) peaks in the near-UV, so Swift observations are required to get a valid measurement of temperature \citep{2022ApJ...937...75A}. Therefore, to assemble single-epoch SEDs, we group our Swift photometry into bins of $\sim$6 hours during the first day of Swift observations and bins of $\sim$1 day thereafter, each of which includes at least one filter with a shorter effective wavelength than $U$. Then we add \textit{BVgri} points by linearly interpolating our high cadence ground-based data. We then fit a blackbody spectrum to each of these single-epoch SEDs using the Light Curve Fitting package \citep{hosseinzadeh_light_2023} to derive a temperature and photospheric radius. These are shown in \autoref{fig:rad-temp}. We find the temperature on +1.2 days to be $24.9 \pm 1.9$ kK, which peaks at $30.7 \pm 1.5$ kK at +1.4 days. The temperatures found here are consistent with temperatures found by \citet{Wynn_2024_ggi} and slightly higher at the peak than \citet{Chen_2024_24ggi}, where they do not include UV data for the temperature calculations. The peak temperature for SN~2024ggi is close to the temperature observed for SN~2023ixf (34.3 kK) by \citet{Zimmerman_2024_2023ixf}. The implied radius at the first epoch is $1571 \pm 156\ R_\sun$ which is smaller than the radius calculated for SN~2023ixf of 2731 $R_\sun$ by \citet{Zimmerman_2024_2023ixf}.

\subsection{Rapid Evolution on the First Day}
In the first $\sim$0.5--1.5 d after explosion, we collected a nearly continuous set of light-curve data, with $\sim$30 data points in each of the $B$, $V$, $g$, $r$, $i$, and Clear/Open bands in that time frame. This early evolution began when the SN was still very faint, with $M_V \approx -12.75$ mag. In addition to the SED fitting described in the previous section, we fit a blackbody SED to each epoch of high-cadence filtered photometry using an MCMC routine implemented in Light Curve Fitting package \citep{hosseinzadeh_light_2023} and then construct a pseudobolometric light curve by integrating the SED between the $U$ and $I$ bands. Since the UV data in earlier phases is limited, we only consider the luminosity between $U$ and $I$ for consistency. This approach facilitates comparisons to previous optical-only data sets. 

\begin{figure}
    \centering
    \includegraphics[width=\columnwidth]{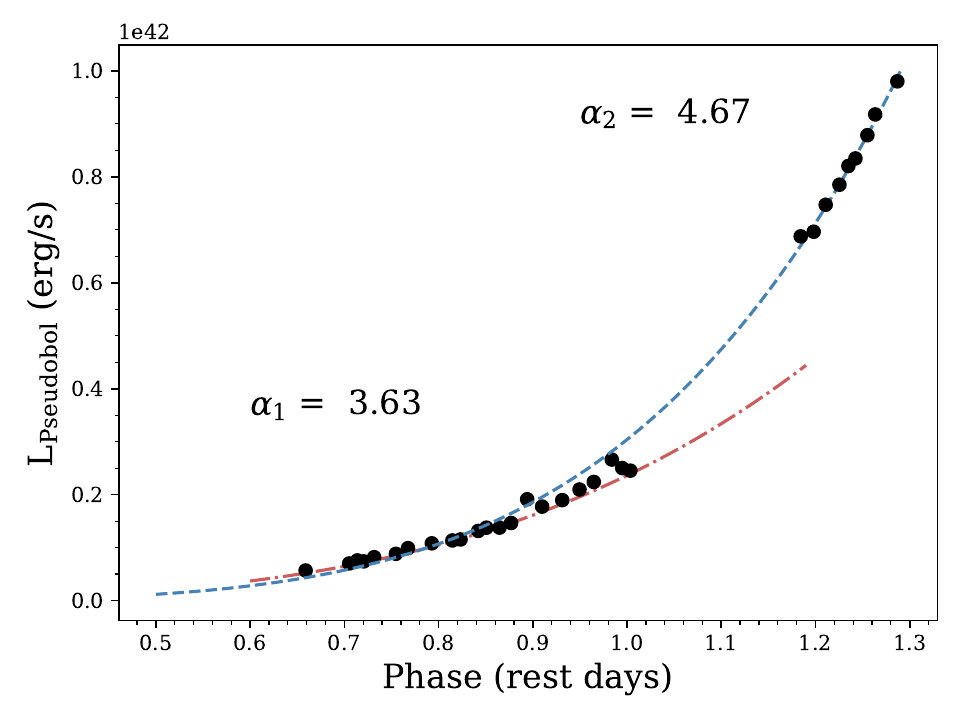} 
    \caption{Pseudobolometric light curve of SN~2024ggi for the first $\lesssim$1.3 days after the explosion. There is a clear evolution in the slope of the luminosity increase in a short period of time. We fit two different power laws to the data set. The very early data is best fit by a power-law index of 3.63 and the second set by a power-law index of 4.67. These fits are shown as solid red and blue lines for each set of data.} 
    \label{fig:bpl}
\end{figure}

When we zoom in on the first day of our pseudobolometric light curve, as shown in Figure~\ref{fig:bpl}, there is a clear change of slope from the first few hours to the end of the first day. We note that we see this change in slope for all the individual filters. This deviation in the early light curve has also been seen in the case of SN~2023ixf (see Figure 2 in \citealt{Hosseinzadeh_2023_23ixf}). We fit a broken power law to this pseudobolometric light curve from +0.6 to +1.3 days. The first set of data spanning up to +1.0 days is best fit by a power law with $\alpha_1 = 3.63$ (red; Figure \ref{fig:bpl}). After that, we see an increase in the slope which is best fit by $\alpha_2 = 4.67$ (blue; Figure \ref{fig:bpl}). This change in power-law index is clearly visible both in the pseudobolometric light curve and individual filters. The initial power-law index is close to the values derived by \citet{Pessi_2024}. However, they do not witness the second part of the light curve rise. This change in slope could be due to the change in the density of CSM where the initial higher density slows down the rise in luminosity and this increases when the density is lower.


\subsection{Color evolution}

\begin{figure}
    \centering
    \includegraphics[width=\columnwidth]{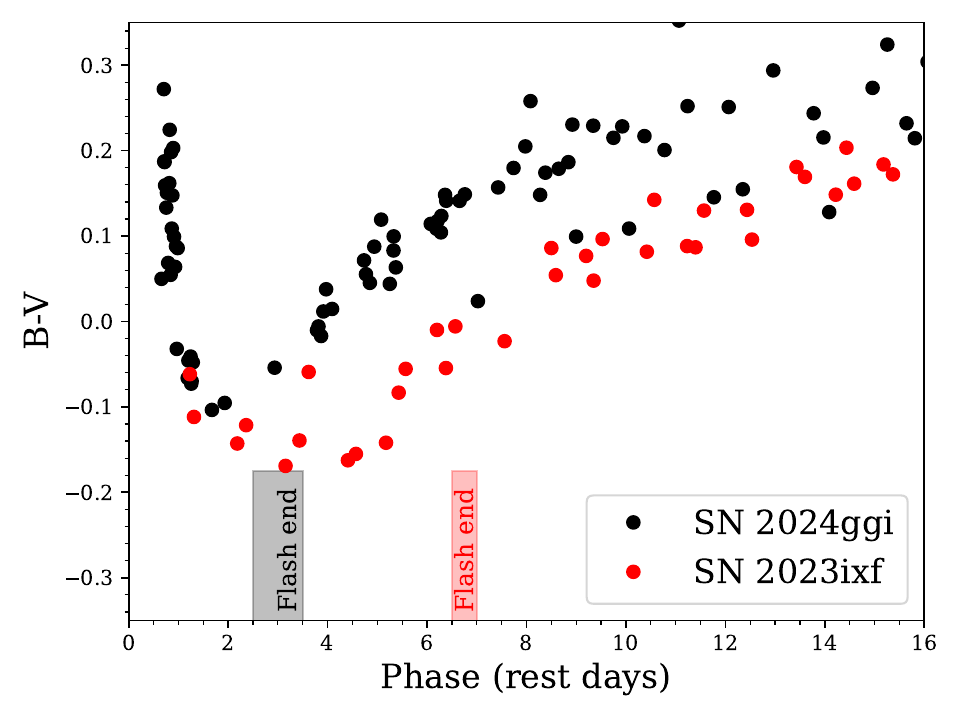}
    \caption{ Extinction corrected B-V color curve of SN~2024ggi (black) and SN~2023ixf (red) for the first 16 days. The last epoch where spectra have flash features for both SN~2024ggi (black) and SN~2023ixf (red) are indicated by the shaded regions. The extent of the shaded region is the length of time between the last spectrum showing a flash feature and the first spectrum with no flash features. 
    The data at phases before 2 days show the color is getting bluer quickly prior to evolving redward.  (The data used to create this figure are available in the published article.)} 
    \label{fig:color-ev}
\end{figure}

We examined the early color evolution of SN~2024ggi with a high-cadence multiwavelength data set. In Figure~\ref{fig:color-ev}, we present the extinction-corrected $B-V$ plot with respect to phase.  During the first few hours, the color gets rapidly bluer. We note that we see similar evolution in other filters such as $r-i$ and $g-r$ as well.
Then at roughly 2 days post-explosion, we reach a blue maximum, after which the color then starts to trend redward. 

We see that the flash features in spectra coincide with blueward color evolution. Once the color starts evolving redward, the narrow emission lines indicative of CSM interaction disappear. Similar color behavior has been seen previously in SN~2023ixf \citep{Hiramatsu_2023,Li_2024_23ixf,Teja_2023_23ixf} and SN~2018zd \citep{Hiramatsu_2021_2018zd}. The turnover in color evolution from blueward to redward happens later in SN~2023ixf than in SN~2024ggi, but this is closer to the time at which flash features disappear for SN~2023ixf (+7 days). Therefore, we attribute the initial blueward color evolution to interaction with dense CSM or delayed/extended shock-breakout in the CSM \citep[e.g.][]{Chugai_2004_sbo, smith_2007_sbo,Ofek_2010_SBO,Dessart_2017_sbo, Haynie_2021_sbo}.
\citet{Hiramatsu_2021_2018zd} likewise interpreted early blueward color evolution in SN~2018zd as a delayed shock breakout through dense CSM, which can provide an additional power source for the explosion and hence a bluer color as the shock front propagates through the CSM.

\subsection{Shock cooling model}\label{sec:lcmodel}
In the absence of significant CSM interaction, the early photometric evolution of SNe II is assumed to be driven by cooling emission from the shock-heated ejecta. A series of increasingly complete prescriptions by \cite{Sapir_2011, Sapir_2013}, \cite{Rabinak_2011}, \cite{Katz_2012}, \cite{Sapir_2017}, and \cite{Morang_2023, 2024MNRAS.528.7137M} model the shock-cooling emission as a function of the properties of the progenitor star and the explosion, in particular the stellar radius. Here, we adopt the model of \citet[hereafter MSW23]{Morang_2023}, which builds on the model of \cite{Sapir_2017} by accounting for the very early ``planar'' phase, where the thickness of the emitting shell is smaller than the stellar radius, and for some line blanketing in UV. This model has been applied to a sample of SNe~II observed by the Zwicky Transient Facility and Swift \citep{2023arXiv231016885I}, as well as several nearby supernovae with very high-cadence early observations \citep{Hosseinzadeh_2023_23ixf,Andrews_2024,Retamal_2024,Shrestha_2024}, with mixed results. These previous works have concluded that the model does not perform well when there is strong circumstellar interaction, which is likely the case here. We note that SN~2024ggi shows signs of CSM interaction but there is value in performing this modeling, which can be used to compare to those cases where SN without CSM interaction is apparent. This can provide crucial knowledge on how CSM interaction affects the light curves and our understanding of shock cooling.

\begin{figure}
    \centering
    \includegraphics[width=\columnwidth]{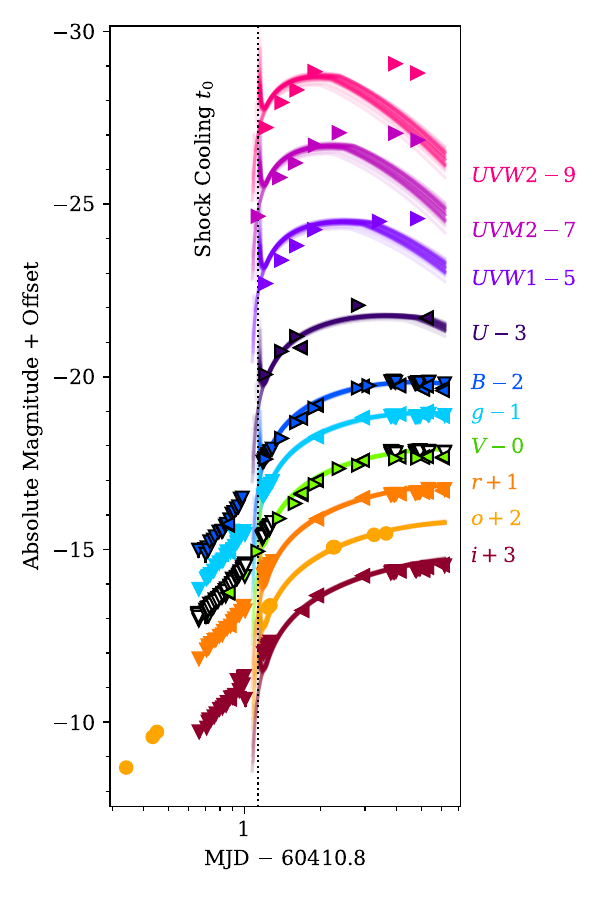}
    \caption{Shock-cooling modeling of SN~2024ggi using the prescription of \citet{Morang_2023}. The model does a good job fitting the rise after the first day of data to the maximum for all filters and the initial plateau for optical filters of SN~2024ggi but does not correctly estimate the explosion time, with SN emission apparent a full day before the model's preferred explosion epoch. Similar behavior and mismatch to the model were seen in the case of  SN~2023ixf \citep{Hosseinzadeh_2023_23ixf}.}
    \label{fig:sc-model}
\end{figure}

We fit the model of MSW23 to the early light curve of SN~2024ggi using a Markov Chain Monte Carlo (MCMC) routine implemented in the Light Curve Fitting package \citep{hosseinzadeh_light_2023}. The priors and best-fit values for each parameter are tabulated in Table~\ref{tab:params}, and the best-fit model is plotted in Figure~\ref{fig:sc-model}. We use observed data from 60411 to 60417 MJD which is the range of time the best-fit model is valid following the prescription described in Appendix of \citet{Morang_2023} (Equation A3). The model could not fit the data from the first 24 hours, which has been seen for SN~2023ixf \citep{Hosseinzadeh_2023_23ixf}. However, for the rest of the data we find that the model converges and gives an overall good fit. 
 The best-fit model gives the explosion epoch to be $60411.947^{+0.008}_{-0.058}$ MJD, which is after the discovery date (60411.65 MJD), due to the code not being able to model the early data points. We also ran the shock cooling model by forcing the upper limit for the explosion epoch to be the discovery date. However, this run was not able to provide a good fit for any part of the light curve. Thus, we pick the model that converges to calculate our parameters. The progenitor radius from the best-fit model is found to be $431^{+14}_{-215}\ R_\sun$. However, we caution that this is not physically representative of the stellar radius due to the extended shock break out observed in the $<$1d light curve, which is unable to be fit by the model.


\begin{deluxetable*}{lCcCCCc}
\label{tab:shockcooling}
\tablecaption{Shock-cooling Parameters\label{tab:params}}
\tablehead{&& \multicolumn{3}{c}{Prior} \\[-10pt]
\colhead{Parameter} & \colhead{Variable} & \multicolumn{3}{c}{------------------------------------------} & \colhead{Best-fit Values\tablenotemark{a}} & \colhead{Units} \\[-10pt]
&& \colhead{Shape} & \colhead{Min.} & \colhead{Max.} }
\startdata
Shock velocity                        & v_\mathrm{s*}  & Uniform & 1      & 5  & 1.7^{+0.4}_{-0.2}            & $10^{8.5}$ cm s$^{-1}$ \\
Envelope mass\tablenotemark{b}        & M_\mathrm{env} & Uniform & 0      & 10    & 0.9^{+0.7}_{-0.3}                    & $M_\sun$ \\
Ejecta mass $\times$ numerical factor & f_\rho M       & Uniform & 0.05    & 100   & 50$\pm$ 40                    & $M_\sun$ \\
Progenitor radius                     & R              & Uniform & 0      & 2000 & 431^{+14}_{-215}& $R_\sun$ \\
Explosion time                        & t_0            & Uniform & 60410.0 & 60413.14  & 60411.947^{+0.008}_{-0.058} & MJD \\
Intrinsic scatter                     & \sigma         & Log-uniform & 0  & 10^2  & 30.1^{+1.1}_{-0.8}                 & \nodata \\
\enddata
\tablenotetext{a}{The ``Best-fit Values'' columns are determined from the 16th, 50th, and 84th percentiles of the posterior distribution, i.e., $\mathrm{median} \pm 1\sigma$. 
}
\end{deluxetable*}
\vspace{-24pt}


\section{Spectroscopy}\label{sec:spec_analysis}
\subsection{Spectral evolution and line identification}
We present a high cadence, low-resolution spectral sequence of SN~2024ggi in Figure~\ref{fig:spec-seq} starting from +0.81 days to +12.6 days with respect to our estimated explosion epoch. The early spectra contain clear flash ionization features indicating the recombination of the CSM ionized by the shock-breakout and ongoing interaction between the ejecta and dense CSM \citep{GalYam_2014,Khazov_2016,Yaron_2017,Tartaglia_2021,Bruch_2021,Hiramatsu_2021_2018zd,Terreran_2022,Bostroem_2023_23ixf,Bruch_2023, Jacobson-Galan_2023, Hiramatsu_2023}. First, we use our SALT RSS spectrum from 60411.979 MJD (+1.18 days) for line identification as it has a higher signal-to-noise ratio than the classification spectrum from the Lijiang 2.4m telescope \citep{Tns_classification_2024}. We clearly detect lines of different ionization levels including \ion{He}{2} $\lambda6559.8$, \ion{C}{3} $\lambda5695.9$,  \ion{C}{4} $\lambda\lambda5801.3$, \ion{N}{3} $\lambda4858.82$, \ion{He}{2} $\lambda4685.5$, \ion{N}{3} $\lambda4640.64$, \ion{N}{3} $\lambda4097.33$, \ion{N}{4} $\lambda \lambda 4057.76$, \ion{O}{5} $\lambda 5597.91$, and~\ion{N}{4} $\lambda\lambda7109$. Compared to the first spectra at +0.81 days, we see a development of high ionization state lines such as \ion{C}{4} and \ion{O}{5} in the +1.18 and +1.19 day spectra. This implies there is an increase in temperature during this phase which is consistent with the temperature and color evolution as shown in \autoref{fig:rad-temp} and \autoref{fig:color-ev} respectively.
These narrow emission features only persist for a few days, and the spectrum from Bok at +3.42 days is almost featureless. The \ion{N}{3} $\lambda4640.64$ line is only visible for $\sim 9$ hours, with the strongest emission seen in the first spectrum. Similarly the \ion{N}{3} $\lambda4097.33$ and \ion{N}{4} $\lambda4057.76$ lines disappear within a day from the first observation. We identify \ion{O}{5} $\lambda 5597.91$, a rare feature in flash spectra, in three different spectra taken at +1.18, +1.19, and +1.75 days, which is consistent with the identification in \citet{Wynn_2024_ggi}.

The rest of the lines persist until +3.42 days. On day +5.48, there is a clear broad absorption feature present in H$\beta$ from which we estimate the ejecta velocity to be ${\sim}9000\ \mathrm{km\,s}^{-1}$ based on the minimum absorption trough. We use this velocity, along with the duration of the flash features (3.42 days) to compute an approximate CSM radius of $R_\mathrm{CSM} \sim 2.7 \times 10^{14} \mathrm{cm}$. This value for the approximate CSM radius is consistent with that of \citet{Wynn_2024_ggi} to within a factor of two. Using the value for $R_\mathrm{CSM}$ and assuming a canonical wind velocity of $v_\mathrm{w} \approx 10~ \mathrm{km\,s^{-1}}$ \citep{Deutsch_1956} we find that the mass loss began $\sim$9 years before explosion.  If instead we assume a velocity of $v_\mathrm{w} = 37\ 
 \pm 4~ \mathrm{km\,s^{-1}} $ derived from the early high-resolution spectra of SN~2024ggi, which is discussed in the next section, we find that the mass loss began just $\sim$2 years before the explosion. Note, we do not see any precursor activity 2 years before the explosion in ATLAS forced photometry data as shown in \autoref{fig:precursor}.

\subsection{High-resolution spectroscopy} \label{sec:highres}
We obtained high-resolution spectra of SN~2024ggi using MIKE on the Magellan-Clay telescope on MJDs 60412.015 and 60412.29, or +1.2 and +1.5 days after the explosion, respectively. In addition, we acquired high-resolution spectra from HRS on SALT on MJDs 60413.96 (+3.16 d), 60419.94 (+9.14 d), and 60419.96 (+9.16 d). We clearly see the \ion{Ca}{2} H \& K and  \ion{Na}{1}~D absorption features in all of our high-resolution spectra both from MIKE and SALT (see Figure~\ref{fig:extinction}). We utilized the equivalent width of \ion{Na}{1}~D1 and \ion{Na}{1}~D2 lines from the host and the Milky Way to calculate the extinction in Section~\ref{sec:extinct}. In addition, we found the host \ion{Na}{1}~D lines are at a redshift of 0.00221, which we adopt as the host redshift for this paper (Table~\ref{tab:results}). Note that this value is slightly offset from the reported redshift of NGC~3621 itself, although it is consistent with the expected rotation curve of the galaxy \citep{Pessi_2024}. 

In our early high-resolution data from MIKE, we identify narrow emission lines associated with the flash ionization which clearly evolve over the $\sim$7 hours between them.

\begin{figure*}
    \centering
    \includegraphics[width=\columnwidth]{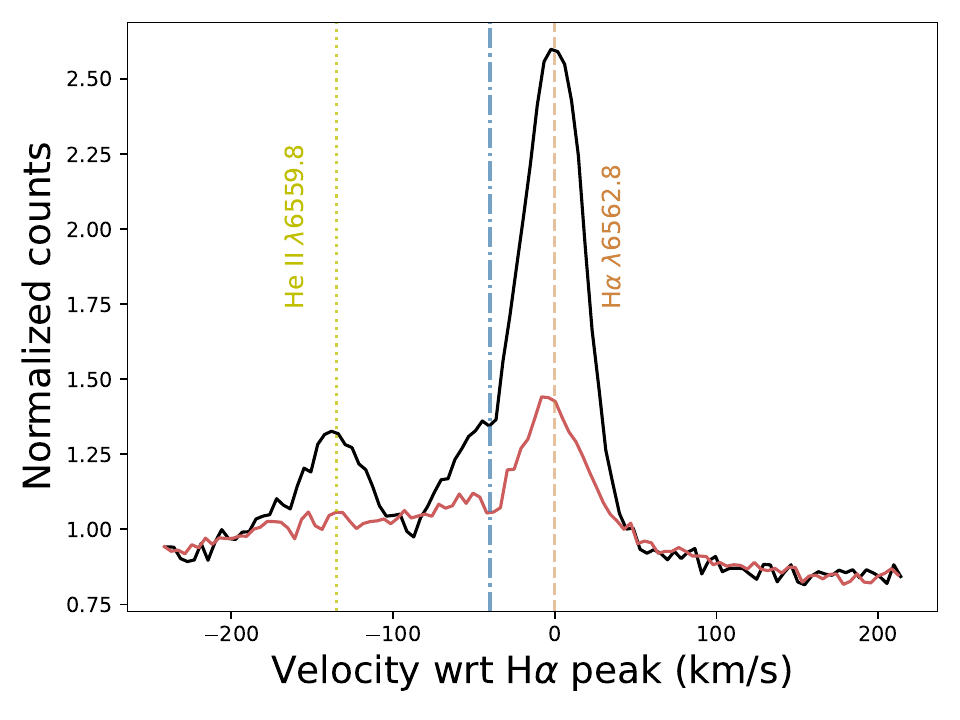}\includegraphics[width=\columnwidth]{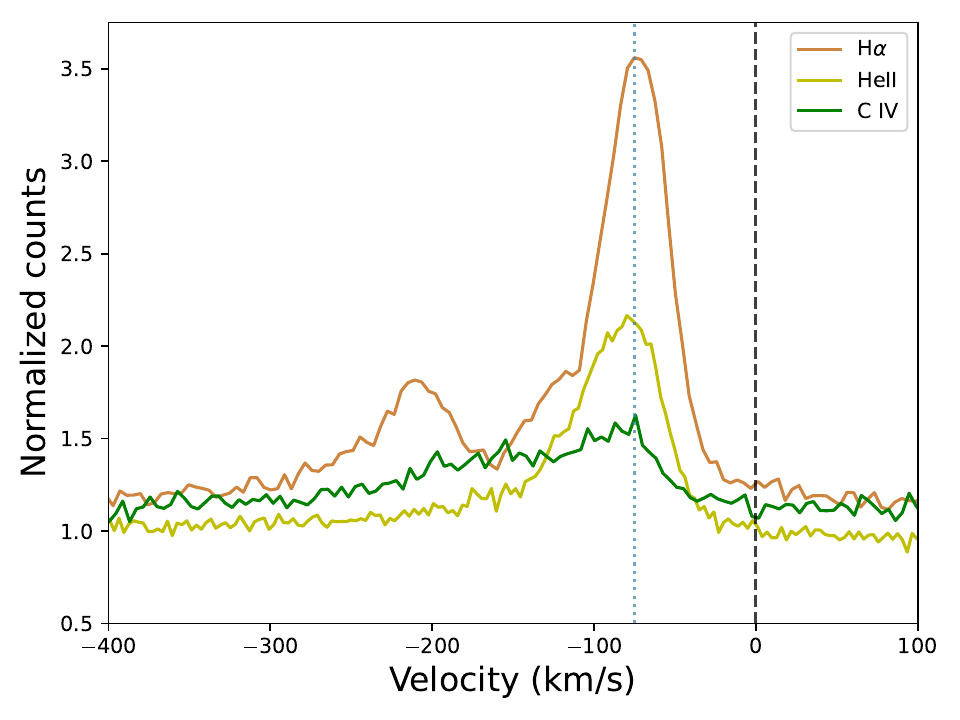}
    \caption{Left: High-resolution MIKE data +1.22 days (black) and +1.49 days (red) focusing on the H$\alpha$ feature. The brown dashed line represents the peak of the feature. We calculate a velocity relative to this wavelength at the position of the shallow absorption feature, a P Cygni signature. The dashed blue line is placed at the shallow absorption feature at $-37\pm 4~ \mathrm{km\,s^{-1}}$, which we use as one of the markers for CSM velocity. \ion{He}{2} is also identified by the yellow dashed line. Right: Line profiles of H$\alpha$, \ion{He}{2} $\lambda 4685.5$, and \ion{C}{4} from the first epoch (+1.22 days), where the black dashed line represents the rest wavelength of these lines. We see a shift in velocity of $-75\ \mathrm{km\,s^{-1}}$ for all of the lines, shown by the blue dotted line.}
    \label{fig:csmV}
\end{figure*}

A close examination of the H$\alpha$ emission line is shown in Figure~\ref{fig:csmV} (left).  The fading in H$\alpha$ is clear.  A weak absorption P Cygni feature is also apparent in both spectra, which we associate with the RSG wind material prior to explosion.  We find the velocity of the absorption feature with respect to the peak of the emission to be at $-37 \pm 4\ \mathrm{km\,s^{-1}}$.  We also fit a Gaussian to the narrow emission line of H$\alpha$  in both MIKE epochs to calculate the full-width-half-maximum (FWHM), which we find to be 36.9 $\mathrm{km\,s^{-1}}$ and 37.0 $\mathrm{km\,s^{-1}}$, respectively.  The velocities from both methods are very similar, and we therefore assume  $37 \pm 4\ \mathrm{km\,s^{-1}} $ as the RSG wind/CSM velocity. We note this value is different from the 79 $\mathrm{km\,s^{-1}}$ calculated by \citet{Pessi_2024}, using the same MIKE data, where they evaluate CSM velocity using the blue shift they observe for the H$\alpha$ feature.

\begin{figure*}
    \centering
    \includegraphics[width = 0.49\textwidth]{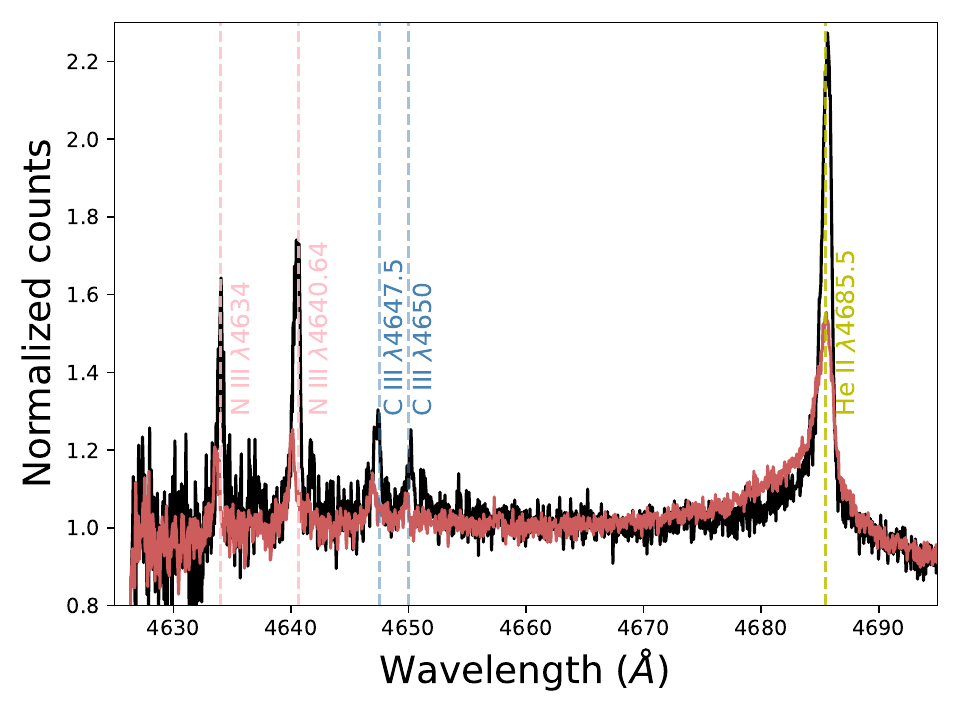} \includegraphics[width = 0.49\textwidth]{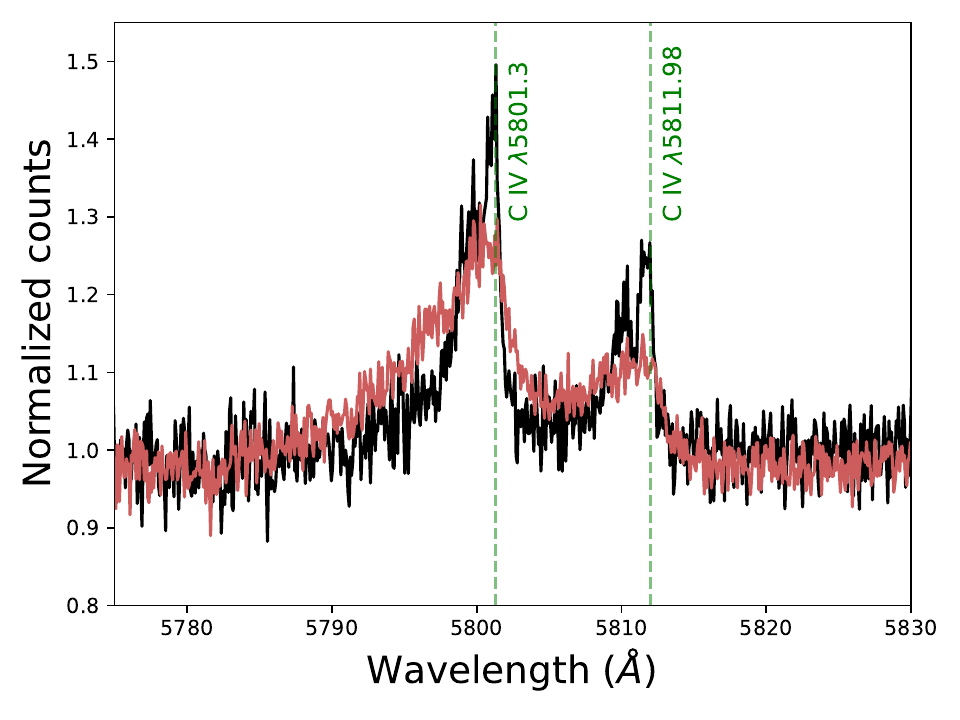} 
    \caption{The two panels show narrow emission lines identified in high-resolution MIKE data from the first (black; +1.22 days) and second (red; +1.49 days) epochs. The spectra are shifted by $-75\ \mathrm{km\,s^{-1}}$ to clearly indicate the narrow emission lines. In the first panel, there are \ion{N}{3}, \ion{C}{3}, and \ion{He}{2} lines identified by pink, blue, and yellow dashed lines, respectively. The second panel zooms in on the \ion{C}{4} lines. 
    The lines evolve rapidly within a few hours between the two epochs.}
    \label{fig:lines}
\end{figure*}

We overall find all the lines in the first two spectra from MIKE to be offset from the rest wavelength by $-75\ \mathrm{km\,s^{-1}}$, after utilizing the \ion{Na}{1}~D-based host redshift as described above (see  Figure~\ref{fig:csmV}, right). In addition, we identify high-ionization lines such as \ion{N}{3}, \ion{C}{3}, \ion{He}{2}, \ion{C}{4}, \ion{He}{2}, and $\mathrm{H \alpha}$ in our first two spectra which are shown in Figure~\ref{fig:lines}. In Figure~\ref{fig:lines}, we have shifted the spectra by $-75\ \mathrm{km\,s^{-1}}$ to align with the rest wavelengths of these lines, shown by the dashed lines. We find that the strengths of the lines change significantly in the 6.6 hours between exposures (see the black and red lines in Figure~\ref{fig:lines}). We measured the equivalent widths of these lines, and they are listed in Table~\ref{tab:lineStrength}. For the \ion{C}{4} line we find the strength increases between the two observation which indicates an increase in ionization and temperature. The rest of the lines show a decrease in strength with time. 

We have a SALT HRS spectrum at +3.16 days, which is towards the end of the phase where we see flash features in low-resolution spectra. Hence, we carefully inspected the high-resolution spectrum looking for any faint and persistent narrow emission lines. We do see a weak, narrow H$\alpha$ line, but due to the low signal-to-noise ratio, we could not perform a quantitative analysis. However, we did not see any other features present in this spectrum. 
We also do not see any narrow emission lines in the SALT HRS spectra at +9.14 and +9.16 days, which agrees well with the behavior seen in low-resolution spectra at similar phases.

\begin{table}
 \caption{Line Strengths from High Resolution Spectra}
 \begin{tabular}{ c c c c }
    \hline
     &  $\lambda$ & \multicolumn{2}{c}{EW (\AA)} \\[-7pt]
    Line &  & \multicolumn{2}{c}{------------------------------------} \\[-7pt]
     &  (\AA) & +1.22 days & +1.49 days \\
    \hline
    H$\alpha$ & 6562.7& 1.61 $\pm$ 0.12  & 0.47 $\pm$ 0.05 \\
    \ion{C}{4} & 5811.98& 1.06 $\pm$ 0.07 & 1.38 $\pm$ 0.09 \\
    \ion{C}{4} & 5801.3& 1.78 $\pm$ 0.09& 2.61 $\pm$ 0.09 \\
    \ion{He}{2} & 4685.5& 1.29 $\pm$ 0.03 &0.49 $\pm$ 0.03 \\
    \ion{C}{3} & 4650& 0.15 $\pm$ 0.04 & 0.04 $\pm$ 0.02 \\
    \ion{C}{3} & 4647.5& 0.21 $\pm$ 0.05 & 0.18 $\pm$ 0.06 \\
   \ion{N}{3} & 4640.64& 0.78 $\pm$ 0.08 & 0.28 $\pm$ 0.11 \\
    \ion{N}{3} & 4634& 0.32 $\pm$ 0.07 & 0.11 $\pm$ 0.01 \\
    
\hline 
    \hline
 \end{tabular}
 
 \label{tab:lineStrength}
\end{table}

\subsection{Comparison with models}

Variations in mass-loss rate, SN luminosity, surface
abundance, and CSM density profile can all affect the
characteristics and evolution of flash-spectroscopy features. Here we compare our low-dispersion spectroscopic data set to two available model grids by
\citet{Dessart_2017} and \citet{Boian_2019} to infer the properties of the progenitor star and its CSM.

\subsubsection{Boian \& Groh}
\begin{figure*}
    \centering
    \includegraphics[width=\textwidth]{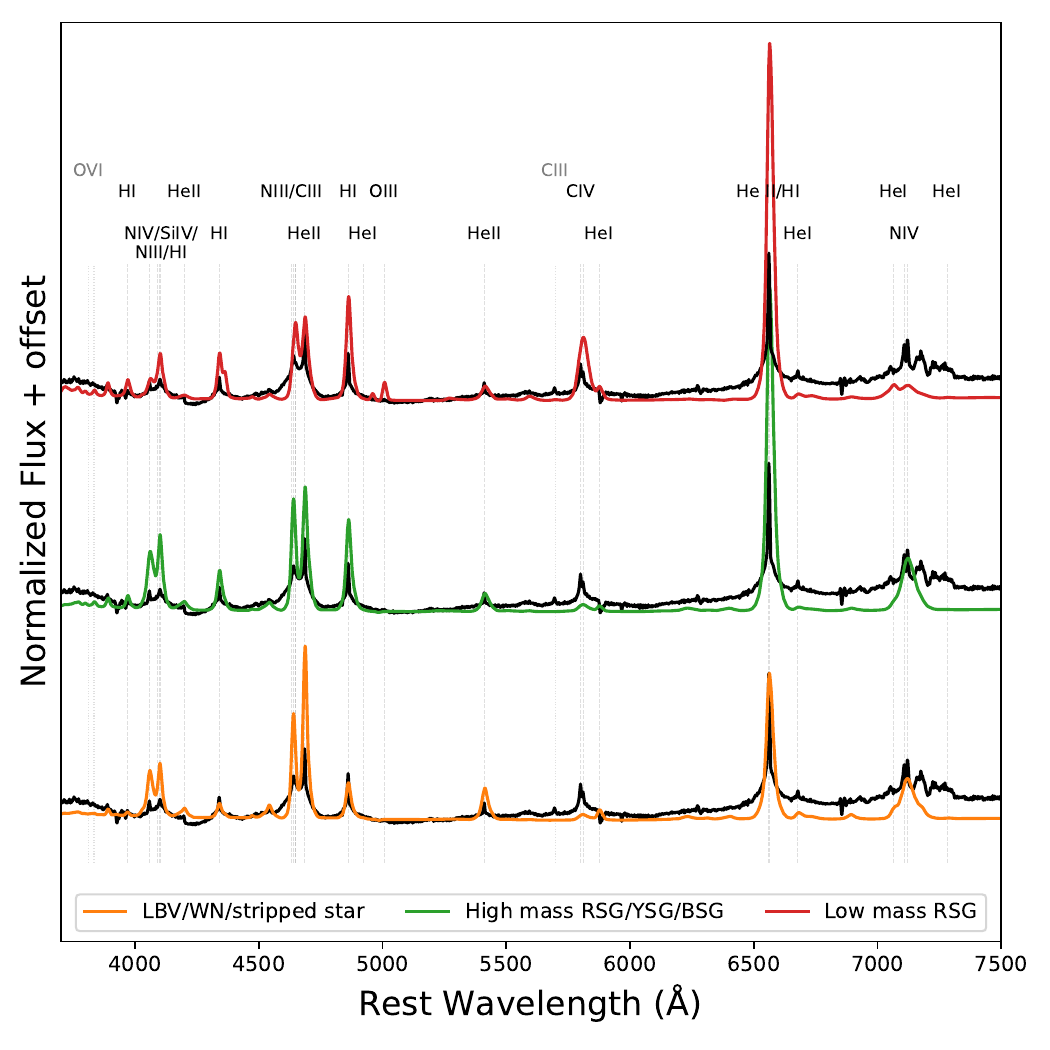}
    \caption{
    Comparison of our +1.27 days spectrum (black) to \citet{Boian_2019} models  with $L = 1.5 \times 10^9\ L_{\sun}$ and $\dot{M} = 3\times 10^{-3}\ M_{\sun}\ {\rm yr}^{-1}$ for three different abundances. The orange line shows the He-rich abundance representing an LBV/WN/stripped star, the green line is for high-mass RSG/BSG/YSG with CNO-processed abundance, and the red is for solar abundance representing a low-mass RSG. Vertical lines represent the ions that are used as a diagnostic of the progenitor in \citet{Boian_2019}. The model spectra have been convolved with a Gaussian kernel to mimic the resolution of the observed spectrum.}
    \label{fig:boin-model}
\end{figure*}

\cite{Boian_2019} used the radiative transfer code CMFGEN \citep{Hillier_1998,hillier_time-dependent_2012,dessart_type_2013,hillier_photometric_2019} to create an extensive spectral library that simulates the interaction of a SN with CSM. They model early spectra approximately one day after explosion with a mass-loss rate ranging from $\dot{M} =  5\times 10^{-4} {-} 10^{-2} \msun{}\,{\rm yr}^{-1}$, SN luminosities $L = 1.9 \times 10^{8} {-} 2.5 \times 10^{10}\ L_{\odot}$, and three different surface abundances corresponding to He-rich abundance representing an LBV/WN/stripped star, CNO-processed abundance for high-mass RSG/BSG/YSG, and the solar abundance representing a low-mass RSG. They make use of the identification of various lines from highly ionized species to low ionization in an early spectrum of SNe interacting with their CSM to map the progenitor properties. These lines are shown in \autoref{fig:boin-model}. We visually inspected the spectral library created by \cite{Boian_2019} to find the most similar spectrum to our earliest SALT RSS spectrum at +1.18 days after the explosion of SN~2024ggi.  

The first SN~2024ggi spectrum observed by SALT at +1.18 days shows flash features including clear lines for \ion{He}{2} ($\lambda4685.5, 4860,5412,6559.8$), \ion{N}{4} ($\lambda4058$, $\lambda\lambda 7109,7122$), \ion{N}{3} ($\lambda\lambda4634, 4640$), \ion{C}{3} ($\lambda\lambda4647, 4650$), and \ion{C}{4} ($\lambda\lambda$ 5801, 5811). For different abundance cases, the only model with clear \ion{N}{3}, \ion{He}{2}, and \ion{N}{4} are seen for $L =1.5 \times 10^{9}\ L_{\odot}$ and $\dot{M} =  3\times 10^{-3}\ \msun{}\ {\rm yr}^{-1}$. The model that matches best with the observed spectrum is shown in Figure~\ref{fig:boin-model} where many of the features match. However, across the full spectrum, none of the observed lines have the right intensity compared to the models. From this model we calculate a mass-loss rate of $\dot{M} =  4\times 10^{-3}\ \msun{}\ \mathrm{yr}^{-1}$, after scaling for the difference in luminosity between the model and the observations, as described in \citet{Boian_2019} and find $L =2.3 \times 10^{9}\ L_{\odot}$ for SN~2024ggi.


\subsubsection{Dessart \& Hillier models}
\begin{figure*}
    \centering
    \includegraphics[width=\textwidth]{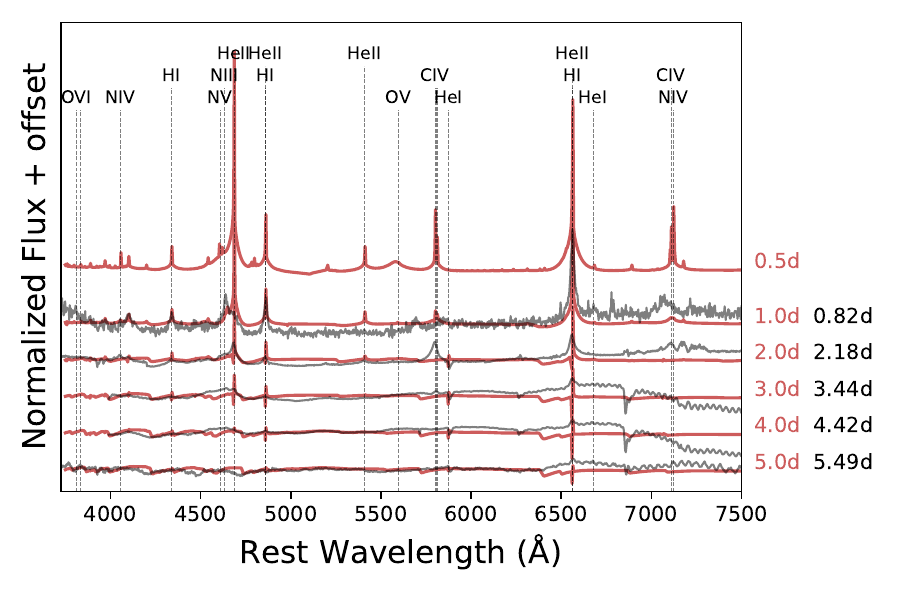} 
    \caption{Comparison of SN~2024ggi spectra (black) with \texttt{r1w4} model (red) from \citet{Dessart_2017}. The observational and model epochs are shown in their respective colors on the right. This is the only model that has the \ion{N}{3} line, which is present in the observed spectrum. Thus, we present this model as the best representation of the observed spectrum. However, the model spectra evolve faster than the observed spectra of SN~2024ggi. This difference in evolution was also seen for SN~2023ixf \citep{Bostroem_2023_23ixf}.  }
    \label{fig:dessart-model}
\end{figure*}

We also visually compared the early spectra of SN~2024ggi with the CMFGEN radiation hydrodynamics model by \citet{Dessart_2017}. These simulations model the spectroscopic signatures of RSG explosions enshrouded in a CSM out to ${\sim} 5 {-} 10\ R_\sun$ and mass ${\leq} 10^{-1}\ M_\sun$. All of the models have the same stellar radius of $501\ R_\sun$ except for \texttt{r2w1}. The mass-loss rate varies from $10^{-6}\ M_\sun{}\ \mathrm{yr}^{-1}$ to $10^{-2}\ M_\sun\ \mathrm{yr}^{-1}$. We visually inspected the simulated spectra from the different models with our data from SALT at +1.18 days after the explosion. In our spectrum, we see flash features described above, which are only present past a few hours in the higher mass-loss rates \citet{Dessart_2017}. We do not find any model that reproduces all of the observed features in the SN~2024ggi spectra. We identify a \ion{N}{3} line next to \ion{He}{2} which is only present for the \texttt{r1w4} case as seen in Figure~\ref{fig:dessart-model}. However, we also identify \ion{C}{4} and \ion{N}{4} lines in the observed spectrum at +1.18 days. These lines disappear by +1 day in the \texttt{r1w4} model while they are present in \texttt{r1w6} until day +3. Thus, we do not find one model that reproduces the observations, but the best fits generally favor a higher-mass-loss model. In addition, the observed spectral evolution diverges significantly from the models starting +2.0 days as the models evolve rapidly and develop absorption features in H$\alpha$, which is not seen in our observed spectra.  The mass-loss rates for \texttt{r1w4} and \texttt{r1w6} are $10^{-3}\ M_\sun{}\ \mathrm{yr}^{-1}$ and $10^{-2}\ M_\sun\ \mathrm{yr}^{-1}$, respectively. This mass-loss-rate range is similar to the value we find from our comparison to the \citet{Boian_2019} model of $\dot{M} =  4 \times 10^{-3}\ \msun{}\ \mathrm{yr}^{-1}$ as shown in Figure~\ref{fig:boin-model}. \citet{Wynn_2024_ggi} compared their spectra with models from \citet{Wynn_2024_model} and found the most compatible model to give a mass loss rate of $10^{-2}\ M_\sun\ \mathrm{yr}^{-1}$. This mass loss rate is comparable to the rate we find by comparing to \citet{Dessart_2017} and \citet{Boian_2019} models.

\section{Comparison with SN~2023ixf}\label{sec:SN23ixf}
\begin{figure}
    \centering
    \includegraphics[width=\columnwidth]{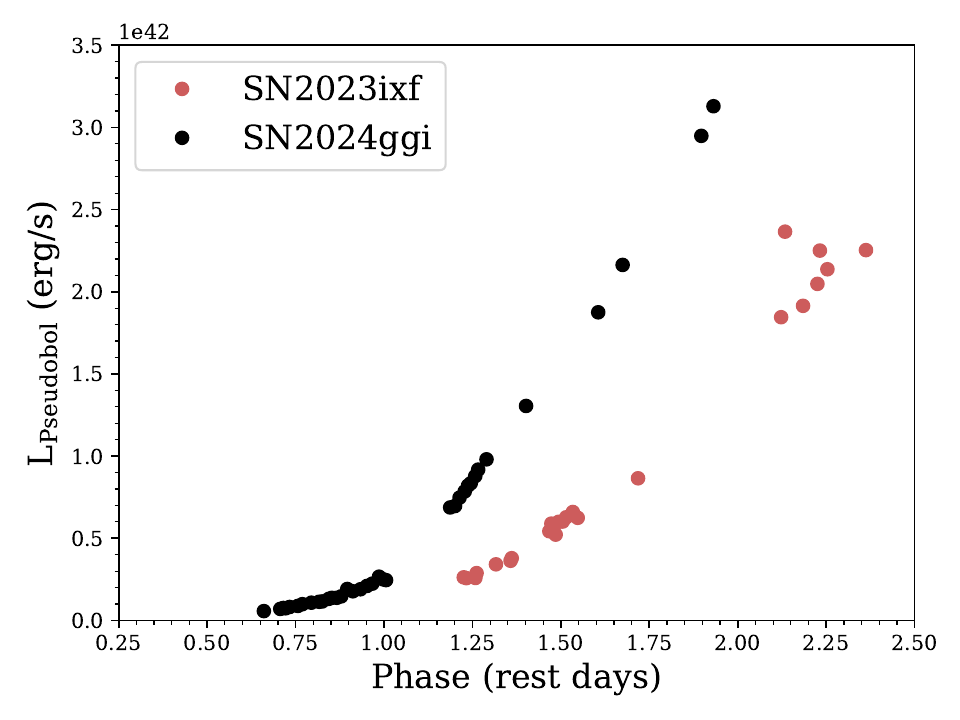}
    \caption{Pseudobolometric light curve of SN~2024ggi (black) compared to SN~2023ixf (red) from \citet{Hosseinzadeh_2023_23ixf}, excluding data from amateur astronomers for the first +2.5 days after the explosion. The rise in luminosity for SN~2024ggi is faster than the rise observed for SN~2023ixf.  }
    \label{fig:flux-comp}
\end{figure}
SN~2023ixf, the recent SN in M101, is a well-studied SN II with flash features at a similar distance to SN~2024ggi. We compare the photometric and spectroscopic behavior of these two SNe in this section. 

First, the light curve in the first 24 hours for both SNe are not described by the simple function $F \propto T^2$, where $F$ is flux and $T$ is time, as might be expected in an ``expanding fireball'' scenario \citep{Hosseinzadeh_2023_23ixf,Hiramatsu2023ApJ...955L...8H}. \citet{Hiramatsu2023ApJ...955L...8H} find that variable mass loss could explain this non-monotonic rise in SN~2023ixf. The shock-cooling model by \citet{Morang_2023} was also not able to match the full rise of the early data before the \citet{itagaki_discovery_2023} discovery, as shown by \citet{Hosseinzadeh_2023_23ixf}. This is similar to what we found for SN~2024ggi. In addition, at comparable phases, the light curve of SN~2024ggi rises faster than that of SN~2023ixf, as shown in Figure~\ref{fig:flux-comp}. This difference could be explained by the details of the CSM density profile and extent between the two SNe, which may be further corroborated by the narrow emission lines of SN~2024ggi disappearing faster than for SN~2023ixf (+3.4 vs +7 days). 
Both SNe show blueward evolution in the very early phase (\autoref{fig:color-ev}), which may be due to the initial shock breakout inside dense CSM. This blueward color evolution roughly coincides with the phases where flash features are seen in the spectra, as shown by the gray and red shaded region in \autoref{fig:color-ev}.
Overall SN~2023ixf remains bluer than SN~2024ggi for the first 16 days, as shown in Figure~\ref{fig:color-ev}. This again could be explained by the CSM interaction lasting longer for SN~2023ixf than for SN~2024ggi.       

\begin{figure*}
    \centering
    \includegraphics[width=\textwidth]{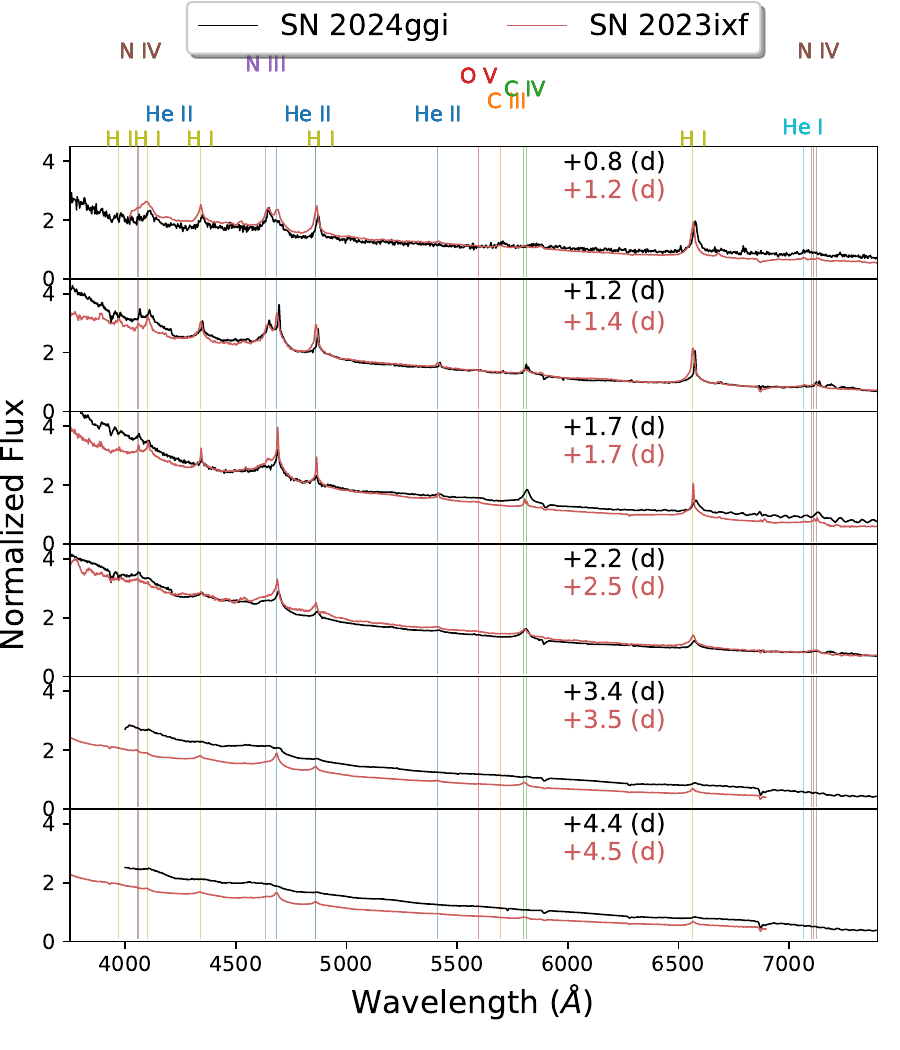}
    \caption{Optical spectral comparison of SN~2024ggi (black) and SN~2023ixf (red) for the first $\sim$4.5 days after explosion. Interesting lines are identified and the epoch for each panel is shown for the two SNe. The first epoch of SN~2023ixf is several hours later than SN~2024ggi, where the \ion{He}{2} line is still developing. Most of the lines are present for both the SNe. However, the strength of the lines rapidly declines for SN~2024ggi, with almost all the flash features vanishing by day 3.5, whereas the SN~2023ixf lines decline more slowly and are present even at +5.3 days. This difference can point to the CSM around SN~2024ggi being more confined than the CSM around SN~2023ixf. }
    \label{fig:comp_23ixf}
\end{figure*}

We compare the spectral evolution of SN~2024ggi with SN~2023ixf as shown in Figure~\ref{fig:comp_23ixf}. We find all the lines present in SN~2023ixf are also in SN~2024ggi. However, in the very first spectrum of SN~2024ggi, we identify \ion{C}{3} $\lambda$ 5695.9 and \ion{O}{5} $\lambda$ 5597.91, which is not seen in SN~2023ixf. We find that the narrow emission lines in SN~2024ggi disappear faster than in SN~2023ixf, and by +3.42 days, all of the narrow emission lines from SN~2024ggi have disappeared, whereas for SN~2023ixf the lines vanish at +7 days \citep{Bostroem_2023_23ixf,Jacobson-Galan_2023,Wynn_2024_ggi,Hiramatsu_2023}. However, for both cases, we get a similar mass loss rate between $10^{-3}$ and $10^{-2}\ M_{\sun}~{\rm yr}^{-1}$ \citep{Bostroem_2023_23ixf} when compared to theoretical models by \citet{Boian_2019} and \citet{Dessart_2017}. Since we find similar mass loss rates for both events, this extended presence of flash features of SN~2023ixf could be due to CSM produced by the progenitor extending farther than the CSM surrounding SN~2024ggi, indicating a different mass-loss history of the progenitor. Even though the mass-loss rates are similar, the smaller CSM radius in SN~2024ggi implies that it has a lower total mass of CSM compared to SN~2023ixf.

From high-resolution spectroscopy, we calculate the CSM velocity of SN~2024ggi to be $37 \pm 4~ \mathrm{km\,s^{-1}}$, which is similar to a typical quiescent RSG wind of 20--30 $\mathrm{km\,s^{-1}}$ \citep{Jura_1990}. Though the calculated CSM velocity is typical for RSG winds, we note that radiative acceleration could contribute to this velocity.
The wind speed of SN~2024ggi is lower than the velocity of 115 $\mathrm{km\,s^{-1}}$ calculated by \citet{Smith_2023_23ixf} for SN~2023ixf where they mention that this higher velocity could be due to the radiation acceleration. Utilizing high-cadence mid-resolution spectroscopy, \cite{2023Scibu..68.2548Z} observed that the velocity of SN~2023ixf increased from 55 $\mathrm{km\,s^{-1}}$ at $t \approx 1.8$ d to 141 $\mathrm{km\,s^{-1}}$ at $t \approx 3.0$ d. Furthermore, \cite{Zhang_2024_24ggi} found that the observed velocity evolution of SN~2023ixf was consistent with a linear growth pattern, confirming that the stellar wind of this SN was accelerated by radiation. Consequently, the upper limit stellar wind velocity of SN~2023ixf before acceleration was 55 $\mathrm{km\,s^{-1}}$, slightly higher than that of SN~2024ggi. This variety in wind velocity shows that we need a larger sample to explore the range of the RSG wind.

In the early high-resolution spectra of SN~2024ggi and SN~2023ixf, a blueshift in all of the emission lines is seen after correcting for the rotational velocity of the host based on \ion{Na}{1}~D absorption lines. However, the amount of blueshift between lines for these SNe is different. For SN~2024ggi, we found all the major lines to be shifted by $-75\ \mathrm{km\,s^{-1}}$ in the first two epochs, whereas, in the case of SN~2023ixf, different lines were shifted by different velocities ranging from 0 to $-150$ or $-200\ \mathrm{km\,s^{-1}}$ \citep{Smith_2023_23ixf}. This blueshift in narrow emission lines has been seen before in the early spectra of interacting SNe, implying that they are emitted by CSM expanding towards the observer relative to the SN \citep{Groh_2014,Shivvers_2015A_198s,Grafener_2016,Smith_2023_23ixf}. It is difficult to compare the velocity evolution of SN 2024ggi to SN~2023ixf, as we have only two high-resolution spectra of SN~2024ggi where flash narrow emission lines are present. 
However, for the first two epochs, we find that the strength of the lines in both SNe evolves in a similar fashion. We find the FWHM of all the lines to decline in the second epoch except for \ion{C}{4}, which is also seen in SN~2023ixf (Figure 4 in \citealt{Smith_2023_23ixf}). 

Finally, for both SNe, no precursor activity is observed, as shown in \autoref{fig:precursor}. The nondetection limit for SN~2023ixf is deeper \citep{Dong2023ApJ...957...28D,Ransome2024ApJ...965...93R} compared to SN~2024ggi. In both cases, precursor activity is not seen during the mass-loss period calculated from $R_\mathrm{CSM}$ and $v_\mathrm{w}$.  


\section{Conclusions} \label{sec:conclude}
We have presented high-cadence, multi-wavelength photometric and spectroscopic follow-up of the nearby Type II SN~2024ggi in NGC3621. In this paper, we focus on the early evolution. The major results from our analysis include:
\begin{enumerate}
     \item We have obtained an unprecedented, high-cadence multi-wavelength photometric data set beginning only a few hours after explosion, providing a great opportunity to study the very early light-curve evolution. We also utilized ATLAS forced photometry at the location of SN~2024ggi to search for precursor activity and found no precursor emission down to $\sim$$-9$ mag, as shown in \autoref{fig:precursor}. 
   \item We find the first 24 hours of data could not be matched by current shock-cooling models (\autoref{fig:sc-model}). This could be due to CSM interaction associated with extended shock breakout. 
   \item The $B-V$ color evolution during this phase shows a unique blueward evolution over the first two days (\autoref{fig:color-ev}). This behavior has been seen previously for very few SNe, and never with the time resolution achieved for SN~2024ggi.
\item In the early spectra of SN~2024ggi, we clearly identify high-ionization lines. The line strengths evolve quickly (over a several-hour time scale), and they vanish by +3.6 days after the explosion. From this, we calculate the CSM radius to be ${\sim}2.3 \times 10^{14}$ cm. 
   \item Comparison of the early spectra with theoretical models by \citet{Boian_2019} and \citet{Dessart_2017} constrain the mass-loss rate for the progenitor star to be $10^{-2} {-} 10^{-3} \ M_{\sun}~ {\rm yr}^{-1}$. 
   \item From high-resolution data we constrain the RSG progenitor wind velocity to be $37 \pm 4\ \mathrm{km\,s^{-1}} $. 
\end{enumerate}


SN~2024ggi has been extensively followed in various wavelengths ranging from $\gamma$-rays to X-rays to radio. There was no detection in $\gamma$-rays, but various groups detected X-ray photons at +2.3, +2.6, and +3.1 days after the explosion \citep{ep_2024_xray,nustar_2024_xray,SGR_2024_xray}. Incidentally, these epochs coincide with the detection of flash features in the optical spectra. In addition, around 3 weeks after the explosion, there was a radio detection at the location of SN~2024ggi \citep{Ryder_2024_24ggiradio}. X-ray and radio detections along with flash features in early spectra all point to the presence of CSM interaction for SN~2024ggi.

The discovery of two core-collapse SNe in the very nearby universe within the last year (SN~2024ggi and SN~2023ixf) provides us with a great opportunity to compare and contrast their properties with high-quality data for years to come. These studies will offer us an unprecedented opportunity to understand the progenitor system, their environment, and the explosion mechanism in greater detail. 
Future observations of SN~2024ggi will add to our understanding of this supernova and could explain the unique early photometric color evolution to the rise of the light-curve, as well as the spectroscopic behavior.

\section*{acknowledgments}
We would like to thank the anonymous referee for their thorough and constructive comments which has helped to improve the paper.

Based on observations obtained at the international Gemini Observatory (GS-2024A-Q-123, PI: Pearson), a program of NSF's NOIRLab, which is managed by the Association of Universities for Research in Astronomy (AURA) under a cooperative agreement with the National Science Foundation. On behalf of the Gemini Observatory partnership: the National Science Foundation (United States), National Research Council (Canada), Agencia Nacional de Investigaci\'{o}n y Desarrollo (Chile), Ministerio de Ciencia, Tecnolog\'{i}a e Innovaci\'{o}n (Argentina), Minist\'{e}rio da Ci\^{e}ncia, Tecnologia, Inova\c{c}\~{o}es e Comunica\c{c}\~{o}es (Brazil), and Korea Astronomy and Space Science Institute (Republic of Korea). 

The SALT spectra presented here were obtained through the Rutgers University SALT programs 2023-1-MLT-008, 2023-2-SCI-030, and 2024-1-MLT-003 (PI: Jha).

Time-domain research by the University of Arizona team and D.J.S.\ is supported by NSF grants AST-1821987, 1813466, 1908972, 2108032, and 2308181, and by the Heising-Simons Foundation under grant \#2020-1864. The research by Y.D., S.V., N.M., and E.H. is supported by NSF grant AST-2008108. 

This work makes use of data from the Las Cumbres Observatory global telescope network.  The LCO team is supported by NSF grants AST-1911225 and AST-1911151.

Research by Y.D., S.V., N.M.R, and E.H. is supported by NSF grant AST-2008108. K.A.B. is supported by an LSSTC Catalyst Fellowship; this publication was thus made possible through the support of Grant 62192 from the John Templeton Foundation to LSSTC. The opinions expressed in this publication are those of the authors and do not necessarily reflect the views of LSSTC or the John Templeton Foundation.  This work makes use of data taken with the Las Cumbres Observatory global telescope network. 
C.P.G. acknowledges financial support from the Secretary of Universities
and Research (Government of Catalonia) and by the Horizon 2020 Research
and Innovation Programme of the European Union under the Marie
Sk\l{}odowska-Curie and the Beatriu de Pin\'os 2021 BP 00168 programme,
from the Spanish Ministerio de Ciencia e Innovaci\'on (MCIN) and the
Agencia Estatal de Investigaci\'on (AEI) 10.13039/501100011033 under the
PID2020-115253GA-I00 HOSTFLOWS project, and the program Unidad de
Excelencia Mar\'ia de Maeztu CEX2020-001058-M. 

This paper includes data gathered with the 6.5m Magellan Telescopes located at Las Campanas Observatory, Chile.

This research has made use of the NASA Astrophysics Data System (ADS) Bibliographic Services, and the NASA/IPAC Infrared Science Archive (IRSA), which is funded by the National Aeronautics and Space Administration and operated by the California Institute of Technology.   This work made use of data supplied by the UK Swift Science Data Centre at the University of Leicester.
GEM acknowledges support from the University of Toronto Arts \& Science Post-doctoral Fellowship program, the Dunlap Institute, and the Natural Sciences and Engineering Research Council of Canada (NSERC) through grant RGPIN-2022-04794.
We acknowledge the support of the staff of the LJT. Funding for the LJT has been provided by the CAS and the People's Government of Yunnan Province. J.Z. is supported by the National Key R\&D Program of China with No. 2021YFA1600404, the National Natural Science Foundation of China (12173082), the Yunnan Province Foundation (202201AT070069), and the International Centre of Supernovae, Yunnan Key Laboratory (No. 202302AN360001).
R.R.M. gratefully acknowledges support by the ANID BASAL project FB210003.

%

\vspace{5mm}
\facilities{Las Cumbres Observatory, SALT (RSS \& HRS), Gemini:South (GMOS), Swift (UVOT),  Magellan:Clay (MIKE), Bok (B\&C)}


\software{Astropy \citep{astropy:2013,astropy:2018, astropy:2022}, Photutils \citep{Bradley_2019}, Panacea, BANZAI \citep{Banzai}, Light Curve Fitting \citep{lightcurvefitting}, Matplotlib \citep{mpl}, Numpy \citep{numpy}, Scipy \citep{scipy}, IRAF \citep{iraf1,iraf2}, PySALT \citep{PySALT},\texttt{lcogtsnpipe}\citep{Valenti_2016}}






\bibliography{SN}{}
\bibliographystyle{aasjournal}



\end{document}